\documentclass[a4paper,showpacs,amsmath,amssymb,12pt]{article}
\usepackage{jheppub} 
\usepackage[T1]{fontenc} 
\usepackage{ulem}
\usepackage{mathrsfs}
\usepackage{amsmath}
\usepackage{amsfonts}
\usepackage{indentfirst}
\usepackage{amssymb}
\usepackage{latexsym}
\usepackage{graphicx}
\usepackage[english]{babel}
\usepackage{epsfig}
\usepackage{subfigure}
\usepackage{multirow}
\usepackage{diagbox}
\usepackage{slashed}
\usepackage[colorlinks=true, linkcolor=blue, citecolor=blue, urlcolor=blue]{hyperref}
\input epsf.sty

\title{\boldmath Further study on excited $\Xi_{QQ^{\prime}}$ via photoproduction at CEPC and FCC-ee}

\author[a,b]{Hong-Hao Ma,}
\author[a,b,1]{Juan-Juan Niu,\note{Corresponding author.}}
\author[c]{Lei Guo}

\affiliation[a]{Department of Physics, Guangxi Normal University,\\Guilin 541004, People's Republic of China}
\affiliation[b]{Guangxi Key Laboratory of Nuclear Physics and Technology, Guangxi Normal University,\\Guilin 541004, People's Republic of China}
\affiliation[c]{Chongqing Key Laboratory for Strongly Coupled Physics, Chongqing University,\\Chongqing 401331, People's Republic of China}

\emailAdd{mahonghao@pku.edu.cn}
\emailAdd{niujj@gxnu.edu.cn}
\emailAdd{guoleicqu@cqu.edu.cn}

\abstract{Within the framework of NRQCD, the photoproduction of doubly heavy baryons $\Xi_{cc}$, $\Xi_{bc}$, $\Xi_{bb}$ and their $P$-wave excited states has been systematically investigated.
The production mechanism is that a color anti-triplet or sextuplet diquark $\langle QQ^{\prime} \rangle$ is first produced, and then evolved into a corresponding doubly heavy baryon $\Xi_{QQ^{\prime}}$ via the subprocess $\gamma+\gamma \rightarrow \langle QQ^{\prime} \rangle[n] +\bar{Q^{\prime}}+\bar{Q} \rightarrow \Xi_{QQ^{\prime}} +\bar{Q^{\prime}}+\bar{Q}$. Here, $Q^{(\prime)}$ denotes the heavy quark $b$ or $c$, [$n$] is the color and spin quantum number of intermediate diquark, which can be $[^3S_1]_{\bar{\textbf{3}}/\textbf{6}}$ and $[^1S_0]_{\bar{\textbf{3}}/\textbf{6}}$ for $S$-wave states, or $[^1P_1]_{\bar{\textbf{3}}/\textbf{6}}$ and
$[^3P_J]_{\bar{\textbf{3}}/\textbf{6}}$ with $J=0,~1,~2$ for $P$-wave states.
Predictions for the cross sections, differential distributions, and theoretical uncertainty have been analyzed. 
The results indicate that, at $\sqrt{s}=91$ GeV, the contribution of photoproduction for $P$-wave $\Xi_{cc}$, $\Xi_{bc}$, and $\Xi_{bb}$ is approximately $2.19\%$, $4.23\%$, $1.26\%$ of the contribution for $S$-wave, respectively. As the collision energy increases, the contribution of $P$-wave also increases. Assuming that the highly excited state can decay into ground state with $100\%$ efficiency, the total produced events at CEPC and FCC-ee can be as high as $\mathcal{O}(10^8)$, $\mathcal{O}(10^7),$ and $\mathcal{O}(10^5)$ corresponding to $\Xi_{cc}$, $\Xi_{bc}$, and $\Xi_{bb}$, respectively, which is very promising to be detected in future experiments.
}

\begin{document}
\maketitle
\flushbottom

\section{Introduction} 

With the enhancement of detectors and the provision of data analysis techniques, more and more hadron states, including mesons, baryons, and exotics, have been discovered experimentally. Baryons are composed of three valence quarks, of which the doubly heavy baryons contain two heavy quarks ($b$ or $c$) and one light quark ($u$, $d$, or $s$), and are widely studied for their rich physical information and complex structures. 
In 2017, the doubly charmed baryon $\Xi_{cc}^{++}$ was first discovered by LHCb Collaboration through the decay channel $\Xi_{cc}^{++} \rightarrow \Lambda_{c}^{+} K^{-} \pi^{+} \pi^{+}$ ($\Lambda_{c}^{+} \rightarrow p K^{-} \pi^{+}$) \cite{LHCb:2017iph}. Subsequently, the LHCb Collaboration confirmed its existence in the decay channel $\Xi_{cc}^{++} \rightarrow \pi^+ \Xi_{c}^+$ \cite{LHCb:2018pcs} and measured the properties of the mass and lifetime of $\Xi_{cc}^{++}$ baryon \cite{LHCb:2018zpl}. In theory, it is devoted to further study of other doubly charmed baryons $\Xi_{cc}$, doubly bottomed baryons $\Xi_{bb}$ and $bc$ baryons $\Xi_{bc}$ predicted by the quark model \cite{GellMann:1964nj,Zweig:1981pd,Zweig:1964jf}, providing certain predictions and guidance for the search of experiments. Many theoretical methods have been proposed to deeply understand the doubly heavy baryons \cite{Berezhnoy:2018bde,Wu:2019gta,Berezhnoy:2018krl,Groote:2017szb,Chen:2018koh,Chen:2019ykv,Sun:2020mvl,Yang:2014tca,Qin:2020zlg,Qin:2021zqx,Ali:2018ifm,Niu:2024ghc,Jiang:2024lsr}, not only the phenomenological extensions of the quark model, but also some nonperturbative methods like the potential model \cite{Bodwin:1994jh,Petrelli:1997ge}, lattice QCD \cite{Kiselev:2000jc}, and QCD sum rules \cite{Kiselev:2002iy}.

One of the main goals of the electron and positron collider is to study properties and interactions of hadron, such as production, decay and mass spectrum. Due to the quark confinement in the strong interactions, the production of doubly heavy baryons contains large and complex non-perturbative effect which cannot be directly obtained by perturbation QCD calculation. As a multi-scale theory, non-relativistic QCD (NRQCD)~\cite{Bodwin:1994jh} provides an efficient means to theoretically study the production of hadrons. Within the framework of NRQCD, the production of hadrons can be factorized into a convolution of two parts, the short-distance coefficient and long-distance matrix elements (LDMEs), based on the perturbative and nonperturbative scale at which the process occurs. In the perturbative region, a diquark can be perturbatively produced with spin and color quantum number $[n]$. Subsequently, the diquark state nonperturbatively transitioned to a certain baryon. 
According to the decomposition of $\rm SU(3)_c$ color group, the color quantum number of diquark can be $\bar{\mathbf 3}$ and $\mathbf 6$. The spin quantum number of diquark can be $[{}^1S_0]$, $[{}^3S_1]$, $[{}^1P_1]$, $[{}^3P_J]$ with $J=0,~1,~2$, and even higher orbital excitation.

At $e^+e^-$ collider, the doubly heavy baryons and their excited states can be produced via direct and indirect production mechanisms, where the direct production mechanism is through the channel mediated by a single virtual photon or $Z^0$ boson \cite{Jiang:2013ej,Niu:2023ojf}, 
$e^+ + e^- \rightarrow \gamma^*/Z^0 \rightarrow \Xi_{QQ^{\prime}} +\bar{Q^{\prime}}+\bar{Q}$, and indirect production mechanism is achieved by the decay of heavy particles \cite{Ma:2022cgt,Niu:2019xuq,Zhang:2022jst,Tian:2023uxe}. However, CEPC provides a potential platform for studying the photoproduction mechanism of doubly heavy baryons, that is, through the subprocess of two photons collision \cite{Zhan:2023jfm,Zhan:2023vwp}, $\gamma+\gamma \rightarrow \Xi_{QQ^{\prime}} +\bar{Q^{\prime}}+\bar{Q}$. The photons in initial-state can be from the bremsstrahlung \cite{Frixione:1993yw} or the laser back-scattering (LBS) \cite{Ginzburg:1981vm} of $e^{+}e^{-}$. The photoproduction mechanism of doubly heavy baryons has also been studied on other platforms, such as LHeC and ILC \cite{Bi:2017nzv,Chen:2014frw}.
Moreover, earlier studies have shown that the results obtained by the photoproduction mechanism are larger than those produced by the direct production mechanism, especially in high energy regions \cite{Chen:2014frw,Zhan:2023jfm}. 

The photoproduction of $P$-wave doubly charmed baryon at future $e^{+} e^{-}$ collider has been investigated in Reference \cite{Zhan:2023vwp}. In this paper, we shall further investigate the photoproduction of $\Xi_{bc}$, and $\Xi_{bb}$ in $P$-wave at Circular Electron Positron Collider (CEPC) \cite{CEPCStudyGroup:2023quu,Ai:2024nmn} and Future Circular Collider (FCC-ee) \cite{FCC:2018evy} within the framework of NRQCD. The integrated cross sections of different intermediate diquark states would be presented under four typical collision energies $\sqrt{s}=$ 91, 160, 240, and 360 GeV of CEPC and FCC-ee.
For the production of $\Xi_{bc}$, the intermediate diquark configurations can be $[{}^1S_0]_{\bar{\textbf{3}}/\textbf{6}}$ and $[{}^3S_1]_{\bar{\textbf{3}}/\textbf{6}}$ for $S$-wave states, $[{}^1P_1]_{\bar{\textbf{3}}/\textbf{6}}$ and
$[{}^3P_J]_{\bar{\textbf{3}}/\textbf{6}}$ with $J=0,~1,~2$ for $P$-wave states; Whereas for $\Xi_{bb}$, because of Fermi-Dirac statistics, there are only half the intermediate states, i.e. $[{}^1S_0]_{\textbf{6}}$ and $[{}^3S_1]_{\bar{\textbf{3}}}$ for $S$-wave states, $[{}^1P_1]_{\bar{\textbf{3}}}$ and
$[{}^3P_J]_{\textbf{6}}$ with $J=0,~1,~2$ for $P$-wave states.
In order to maintain the integrity of the doubly heavy baryon $\Xi_{QQ^{\prime}}$ and facilitate comparative analysis, the results of $\Xi_{cc}$ are also presented in numerical analysis. The differential distributions involving transverse momentum, rapidity, invariant mass and angular distributions, as well as the theoretical uncertainty of the doubly heavy baryons at the collision energy of $\sqrt{s}=91$ GeV are also analyzed accordingly, and the cross section will be suppressed as the collision energy increases.

The rest of this paper is arranged as follows. In Section \ref{sec2}, we present the calculation technology
for the photoproduction of excited doubly heavy baryons, $\gamma+\gamma \rightarrow \Xi_{QQ^{\prime}} +\bar{Q^{\prime}}+\bar{Q}$. The numerical results of $\Xi_{cc}$, $\Xi_{bc}$, and $\Xi_{bb}$ in $S$-wave and $P$-wave are listed in Section \ref{sec3}. Finally, Section \ref{sec4} is reserved for a summary.

\section{Calculation technology}\label{sec2}

Within the framework of NRQCD, the cross section for the photoproduction of excited doubly heavy baryons, $\gamma+\gamma \rightarrow \Xi_{QQ^{\prime}} +\bar{Q^{\prime}}+\bar{Q}$, can be factorized into

\begin{eqnarray}
	\mathrm{d} \sigma(e^{+} e^{-} \rightarrow  \Xi_{QQ^{\prime}}+\bar{Q^{\prime}}+\bar{Q})
	= \int \mathrm{d} x_{1} f_{\gamma / e}(x_{1}) \int \mathrm{d} x_{2} f_{\gamma / e}(x_{2})\nonumber\\
	 \times  \sum_{n} \mathrm{~d} \hat{\sigma}(\gamma\gamma \rightarrow \langle QQ^{\prime}\rangle [n]+\bar{Q^{\prime}}+\bar{Q})\left\langle \mathcal{O}^{H}[n]\right\rangle,
\end{eqnarray}
where $Q$ and $Q^{\prime}$ stand for the heavy $c$- or $b$-quark for the production of $\Xi_{cc}$, $\Xi_{bc}$, and $\Xi_{bb}$, respectively; $f_{\gamma/e}(x)$ represents the energy spectrum of the photon;
$\mathrm{~d} \hat{\sigma}(\gamma\gamma \rightarrow \langle QQ^{\prime}\rangle [n]+\bar{Q^{\prime}}+\bar{Q})$ denotes the differential partonic cross section, which needs to be summed over the color and spin quantum number $[n]$ of the diquark state $\langle QQ^{\prime}\rangle$. It combined with the photon energy spectrum can be called the perturbative short-distance coefficient. $\langle{\cal O}^{H}[n]\rangle$ is the nonperturbative long-distance matrix elements representing the hadronization from the perturbative diquark $\langle QQ^{\prime}\rangle[n]$ to the corresponding baryon $\Xi_{QQ^{\prime}}$.

\subsection{Short-distance coefficient}

\begin{figure}
\centering
 \subfigure[]{
    \includegraphics[scale=0.25]{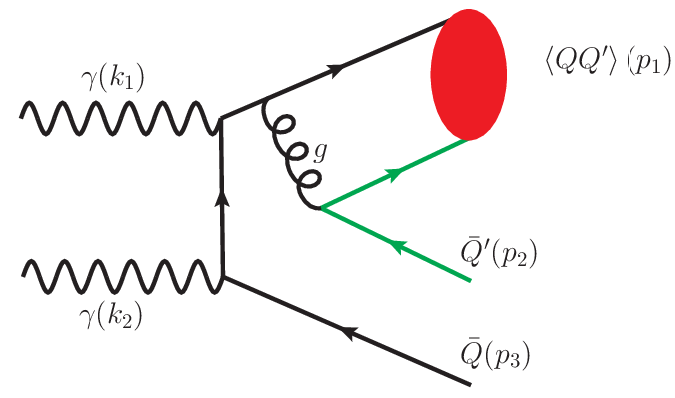}}
  \subfigure[]{
    \includegraphics[scale=0.25]{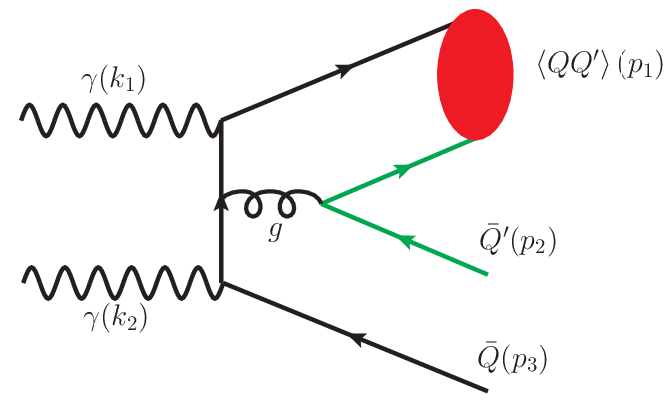}}
  \subfigure[]{
    \includegraphics[scale=0.22]{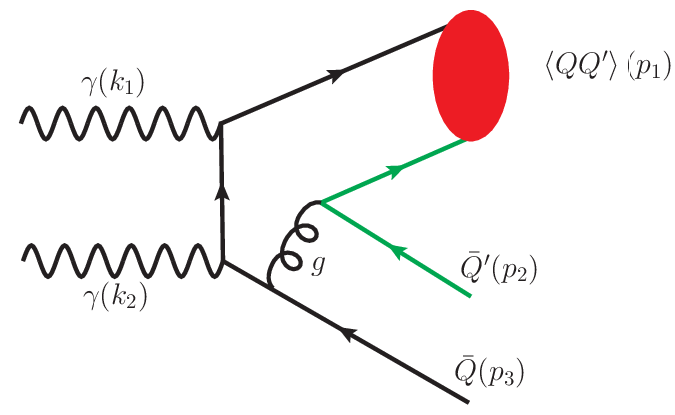}}
  \subfigure[]{   
    \includegraphics[scale=0.22]{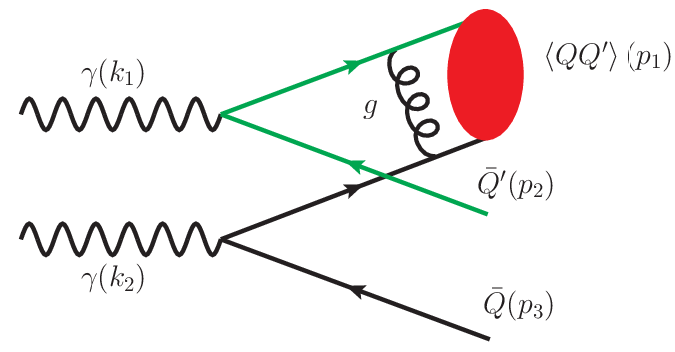}}
  \subfigure[]{
    \includegraphics[scale=0.22]{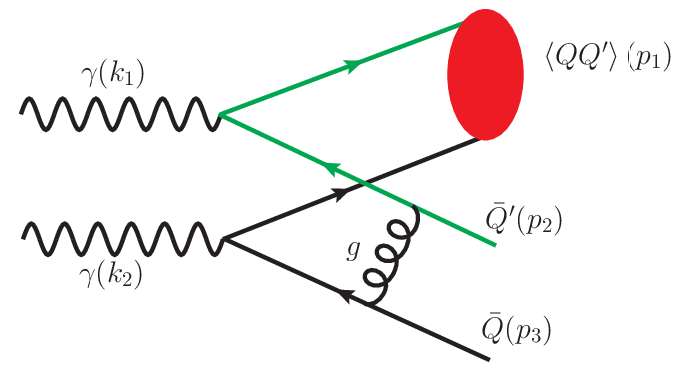}}\\
  \subfigure[]{
    \includegraphics[scale=0.25]{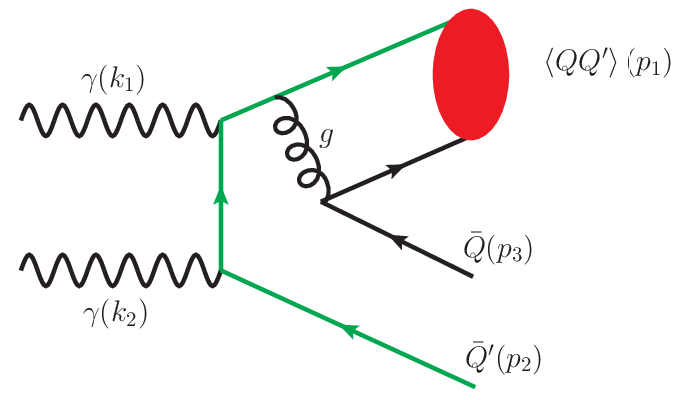}}
  \subfigure[]{  
    \includegraphics[scale=0.25]{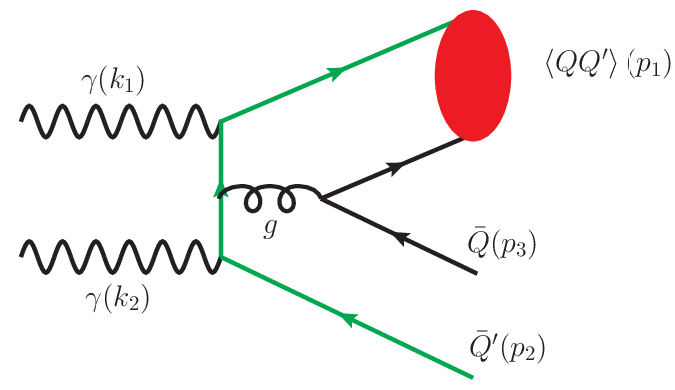}}
  \subfigure[]{ 
    \includegraphics[scale=0.22]{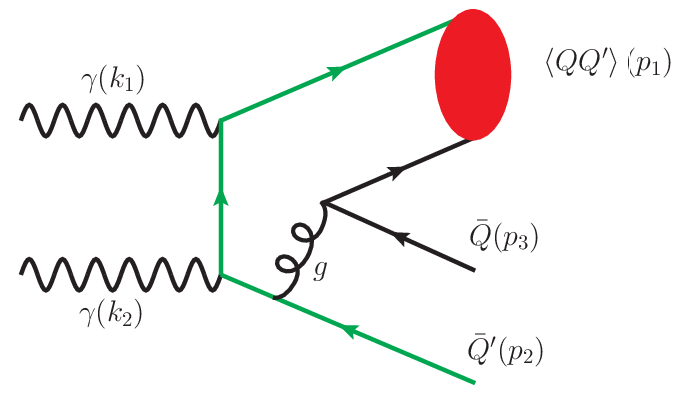}}
  \subfigure[]{
    \includegraphics[scale=0.22]{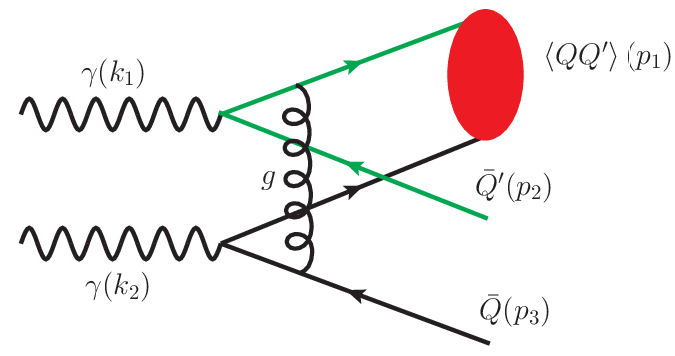}}
  \subfigure[]{
    \includegraphics[scale=0.22]{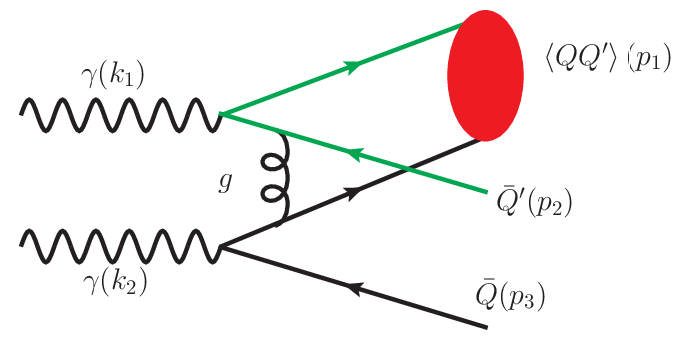}}
\caption{Typical Feynman diagrams for the photoproduction of $\Xi_{QQ^{\prime}}$.} 
\label{fm}
\end{figure}

The typical Feynman diagrams of the partonic processes for $\Xi_{QQ^{\prime}}$ photoproduction are plotted in Fig.~\ref{fm} by JaxoDraw~\cite{Binosi:2003yf}.

The energy distribution of initial photons from laser back-scattering (LBS) of $e^+e^-$ has been well delineated in the spectrum by Ginzburg et al.~\cite{Ginzburg:1981vm},
\begin{equation}
	f_{\gamma/e}(x)=\frac{1}{N}\left[1-x+\frac{1}{1-x}-4 r(1-r)\right],
\end{equation}
where $x=E_{\gamma} / E_{e}$, $r=x /(x_{m}(1-x))$, and the normalization factor
	$N=(1-\frac{4}{x_{m}}-\frac{8}{x_{m}^{2}}) \log(1+x_m)+\frac{1}{2}+\frac{8}{x_{m}}-\frac{1}{2 (1+x_m)^{2}}$.
Here, $x_{m}=4 E_{e} E_{l} \cos ^{2} \frac{\theta}{2}$ with the energy of the incident electron and laser beams $E_e$ and $E_l$, and their angle $\theta$. The range of LBS photon energy is constrained by
\begin{eqnarray}
0 \leq x \leq \frac{x_m}{1+x_m},
\end{eqnarray}

where ( $x_m=4.83$~\cite{Telnov:1989sd}). The differential partonic cross section $\mathrm{~d} \hat{\sigma}$ can be rewritten as

\begin{eqnarray}
d \hat{\sigma}(\gamma\gamma \rightarrow \langle QQ^{\prime}\rangle [n]+\bar{Q^{\prime}}+\bar{Q})=\frac{1}{2x_1x_2s}\overline{\sum}|\mathcal{M}[n]|^2 d\Phi_3,
\end{eqnarray}
where $\sqrt{s}$ is the collision energy of $e^+e^-$ collider (CEPC or FCC-ee); $\overline{\sum}$
means to average over the spin states of the photons, and sum over the color and spin of the final-state
particles; the quantum number $[n]$ in the magnitude $\mathcal{M}$ can be $[{}^1S_0]_{\textbf{6}}$,
$[{}^3S_1]_{\bar{\textbf{3}}}$, $[{}^1P_1]_{\bar{\textbf{3}}}$ and $[{}^3P_J]_{\textbf{6}}$ with $J=0, 1, 2$ for $\Xi_{cc}$ and $\Xi_{bb}$ baryons. While for $\Xi_{bc}$ baryon, there are twelve intermediate diquark states, involving $[{}^1S_0]_{\bar{\textbf{3}}/\textbf{6}}$, $[{}^3S_1]_{\bar{\textbf{3}}/\textbf{6}}$, $[{}^1P_1]_{\bar{\textbf{3}}/\textbf{6}}$ and $[{}^3P_J]_{\bar{\textbf{3}}/\textbf{6}}$ with $J=0, 1, 2$; the three-body phase space
\begin{eqnarray}
d \Phi_3=(2\pi)^4\delta^4(k_1+k_2-\sum_{f=1}^3 p_f)\prod_{f=1}^3 \frac{d\vec{p}_f}{(2\pi)^3 2p_f^0}.
\end{eqnarray}

In leading order, there are a total of twenty Feynman diagrams for the photoproduction $\Xi_{QQ^{\prime}}$, among which ten typical diagrams are shown in Fig.~\ref{fm}, and another ten diagrams can be obtained by exchanging the initial-state photons. However, for $\Xi_{cc}$ and $\Xi_{bb}$ baryons, twenty more diagrams can be produced by swapping the identical $c$-quark lines in the diquark. Therefore, an additional factor $\frac{2^2}{2!2!}$ should be multiplied, where $2^2$  contributes to the twenty additional Feynman diagrams, and 1/(2!2!) is the identical factor of
two heavy quarks inside the diquark and two final-state antiquarks. The momenta of the constituent quark in diquark $p_{11}=\frac{m_Q}{m_{QQ^{\prime}}}p_1+p$ and $p_{12}=\frac{m_{Q^{\prime}}}{m_{QQ^{\prime}}}p_1-p$ with the small relative momentum $p$ between two constituent quarks and $p=0$ in the $S$-wave amplitude.

To obtain the amplitude, one of the fermion lines in the Feynman diagrams (green in Fig.~\ref{fm}) needs to be reversed by applying the charge conjugate matrix $C=-i \gamma^2 \gamma^0$. The action of $C$ parity obeys
\begin{eqnarray}
&&v_{s_2}^T(p_{2}) C  =-\bar{u}_{s_2}(p_{2}), 
~~~~~~~~~~C^{-} \bar{u}_{s_1}(p_{12})^T =v_{s_1}(p_{12}), \nonumber \\
&&C^{-} (\gamma^\mu)^T C =-\gamma^\mu,~~~~~~~~~~~~~C^{-} (\gamma^\mu \gamma^5)^T C =\gamma^\mu \gamma^5,\\
&&C^{-} S_F^T(-q_i, m_i) C  =S_F(q_i, m_i).\nonumber
\end{eqnarray}

Remarkably, after the action of $C$ parity, there is an additional factor $(-1)^{\zeta+1}$, where $\zeta$ is the number of vector vertices in the reversed $Q^{\prime}\bar{Q^{\prime}}$ fermion chain. In this paper, $\zeta=1,$ $2$, or $3$ for the photoproduction of $\Xi_{QQ^{\prime}}$. Let's take a general fermion line as an example to explain the action of charge conjugation, i.e., $L_1=\bar{u}_{s_1}(p_{12}) \Gamma_{i+1} S_F(q_i, m_i) \cdots S_F(q_1, m_1) \Gamma_1v_{s_2}(p_2)$. 
Here, $\Gamma_{i+1}$ represents the interaction vertex, $S_F(q_i, m_i)$ denotes the fermion propagator, where $q_i$ and $m_i$ are the corresponding momentum and mass with the index $i$ of the number of fermion propagators $(i=0,1, \ldots)$ along this fermion line. The subscripts $s_1$ and $s_2$ are the spin indices of the final-state quark and antiquark. With the action of $C$ parity, $L_1$ can be reversed to
\begin{equation}
	\begin{aligned}
		L_1 & =L_1^T=v_{s_2}^T(p_2) \Gamma_1^T S_F^T(q_1, m_1) \cdots S_F^T(q_i, m_i) \Gamma_{i+1}^T \bar{u}_{s_1}^T(p_{12}) \\
		& =v_{s_2}^T(k_2) C C^{-} \Gamma_1^T C C^{-} S_F^T(q_1, m_1) C C^{-} \cdots C C^{-} S_F^T(q_i, m_i) C C^{-} \Gamma_{i+1}^T C C^{-} \bar{u}_{s_1}^T(p_{12}) \\
		& =(-1)^{(\zeta+1)} \bar{u}_{s_2}(p_2) \Gamma_1 S_F(-q_1, m_1) \cdots S_F(-q_i, m_i) \Gamma_{i+1} v_{s_1}(p_{12}) ,
	\end{aligned}
\end{equation}

Specifically, the $S$-wave amplitude $\mathcal{M}[n]=\sum_{i=1}^{20}\mathcal{M}_i[n]$, ten of these can be
written in sequence according to the Feynman diagrams in Fig. \ref{fm} and are listed in Eqs.~\ref{eq10}. The remaining ten diagrams can be obtained by exchanging the positions of the initial photons.

\begin{eqnarray}
{\cal M}_{1}[n] &=& {\cal C} Q_e^{2} \bar{u}_{s'}(p_2) \gamma^{\sigma} \frac{\Pi[n](p_1)}{(p_2+p_{12})^2} \gamma_{\sigma} \frac{{\not\!p}_1+{\not\!p}_2+m_Q}{(p_1+p_2)^2-m_Q^2} \not\!\varepsilon(k_1) \frac{{\not\!k}_2-{\not\!p}_3+m_Q}{(k_2-p_3)^2-m_Q^2} \not\!\varepsilon(k_2) v_{s}(p_3),  \label{eq1} \nonumber\\
{\cal M}_{2}[n] &=& {\cal C} Q_e^{2} \bar{u}_{s'}(p_2) \gamma^{\sigma} \frac{\Pi[n](p_1)}{(p_2+p_{12})^2} \not\!\varepsilon(k_1) \frac{{\not\!p}_{11}-{\not\!k}_1+m_Q}{(p_{11}-k_1)^2-m_Q^2} \gamma_{\sigma} \frac{{\not\!k}_2-{\not\!p}_3+m_Q}{(k_2-p_3)^2-m_Q^2} \not\!\varepsilon(k_2) v_{s}(p_3),  \nonumber\\
{\cal M}_{3}[n] &=& {\cal C} Q_e^{2} \bar{u}_{s'}(p_2) \gamma^{\sigma} \frac{\Pi[n](p_1)}{(p_2+p_{12})^2} \not\!\varepsilon(k_1) \frac{{\not\!p}_{11}-{\not\!k}_1+m_Q}{(p_{11}-k_1)^2-m_Q^2} \not\!\varepsilon(k_2) \frac{-{\not\!p}_3-{\not\!p}_2-{\not\!p}_{12}+m_Q}{(p_3+p_2+p_{12})^2-m_Q^2} \gamma_{\sigma} v_{s}(p_3),  \nonumber\\
{\cal M}_{4}[n] &=& -{\cal C} Q_e Q_e^{\prime} \bar{u}_{s'}(p_2) \not\!\varepsilon(k_1) \frac{{\not\!p}_2-{\not\!k}_1+m_{Q'}}{(p_2-k_1)^2-m_{Q'}^2} \gamma^{\sigma} \frac{\Pi[n](p_1)}{(k_2-p_3-p_{11})^2} \gamma_{\sigma} \frac{{\not\!k}_2-{\not\!p}_3+m_{Q}}{(k_2-p_3)^2-m_{Q}^2} \not\!\varepsilon(k_2) v_{s}(p_3),  \nonumber\\
{\cal M}_{5}[n] &=& -{\cal C} Q_e Q_e^{\prime} \bar{u}_{s'}(p_2) \gamma^{\sigma} \frac{{\not\!k}_1-{\not\!p}_{12}+m_{Q'}}{(k_1-p_{12})^2-m_{Q'}^2} \not\!\varepsilon(k_1) \frac{\Pi[n](p_1)}{(k_1-p_2-p_{12})^2} \not\!\varepsilon(k_2) \frac{{\not\!p}_{11}-{\not\!k}_2+m_{Q}}{(p_{11}-k_2)^2-m_{Q}^2} \gamma_{\sigma} v_{s}(p_3), \nonumber\\
{\cal M}_{6}[n] &=& {\cal C} Q_e^{\prime 2} \bar{u}_{s'}(p_2) \not\!\varepsilon(k_2) \frac{{\not\!p}_2-{\not\!k}_2+m_{Q'}}{(p_2-k_2)^2-m_{Q'}^2} \not\!\varepsilon(k_1) \frac{-{\not\!p}_1-{\not\!p}_3+m_{Q'}}{(p_1+p_3)^2-m_{Q'}^2} \gamma^{\sigma} \frac{\Pi[n](p_1)}{(p_3+p_{11})^2} \gamma_{\sigma} v_{s}(p_3), \nonumber\\
{\cal M}_{7}[n] &=& {\cal C} Q_e^{\prime 2} \bar{u}_{s'}(p_2) \not\!\varepsilon(k_2) \frac{{\not\!p}_2-{\not\!k}_2+m_{Q'}}{(p_2-k_2)^2-m_{Q'}^2} \gamma^{\sigma}  \frac{{\not\!k}_1-{\not\!p}_{12}+m_{Q'}}{(k_1-p_{12})^2-m_{Q'}^2} \not\!\varepsilon(k_1) \frac{\Pi[n](p_1)}{(p_3+p_{11})^2} \gamma_{\sigma} v_{s}(p_3),  \nonumber\\
{\cal M}_{8}[n] &=& {\cal C} Q_e^{\prime 2} \bar{u}_{s'}(p_2) \gamma^{\sigma} \frac{{\not\!p}_3+{\not\!p}_2+{\not\!p}_{11}+m_{Q'}}{(p_3+p_2+p_{11})^2-m_{Q'}^2} \not\!\varepsilon(k_2) \frac{{\not\!k}_1-{\not\!p}_{12}+m_{Q'}}{(k_1-p_{12})^2-m_{Q'}^2} \not\!\varepsilon(k_1) \frac{\Pi[n](p_1)}{(p_3+p_{11})^2} \gamma_{\sigma} v_{s}(p_3), \nonumber\\
{\cal M}_{9}[n] &=& -{\cal C} Q_e Q_e^{\prime} \bar{u}_{s'}(p_2) \not\!\varepsilon(k_1) \frac{{\not\!p}_2-{\not\!k}_1+m_{Q'}}{(p_2-k_1)^2-m_{Q'}^2} \gamma^{\sigma} \frac{\Pi[n](p_1)}{(p_2-k_1-p_{12})^2} \not\!\varepsilon(k_2) \frac{{\not\!p}_{11}-{\not\!k}_2+m_{Q}}{(p_{11}-k_2)^2-m_{Q}^2} \gamma_{\sigma} v_{s}(p_3), \nonumber\\
{\cal M}_{10}[n] &=& -{\cal C} Q_e Q_e^{\prime} \bar{u}_{s'}(p_2) \gamma^{\sigma} \frac{{\not\!k}_1-{\not\!p}_{12}+m_{Q'}}{(k_1-p_{12})^2-m_{Q'}^2} \not\!\varepsilon(k_1) \frac{\Pi[n](p_1)}{(k_2-p_3-p_{11})^2} \gamma_{\sigma} \frac{{\not\!k}_2-{\not\!p}_3+m_{Q}}{(k_2-p_3)^2-m_{Q}^2} \not\!\varepsilon(k_2) v_{s}(p_3). \nonumber\\
\label{eq10}
\end{eqnarray}

Here, $Q^{(\prime)}_e = \frac{2}{3}$ or $-\frac{1}{3}$ for $c$- or $b$-quark, respectively, $\varepsilon(k_1)$ and $\varepsilon(k_2)$ are the polarization vectors of the initial photons. ${\cal C}$ is the overall constant and satisfies ${\cal C}=\mathcal{C}_{i j, k}e^2 g_{s}^2$. Note that in order to obtain the amplitude of the diquark state, the spinor $v_{s {12}}(p_{12}) \bar{u}_{s {11}}(p_{11})$ in the amplitudes has been replaced by the spin projector $\Pi[n](p_1)$, of the following form 

\begin{eqnarray}
\Pi[^1 S_0](p_1) &=& \frac{-\sqrt{m_{QQ^{\prime}}}}{4 m_{Q} m_{Q^{\prime}}}\left(\slashed{p}_{12}-m_{Q^{\prime}}\right) \gamma^{5}\left(\slashed{p}_{11}+m_{Q}\right),\\
\Pi[^3 S_1]^{\beta}(p_1)&=& \frac{-\sqrt{m_{QQ^{\prime}}}}{4 m_{Q} m_{Q^{\prime}}}\left(\slashed{p}_{12}-m_{Q^{\prime}}\right) \gamma^{\beta} \left(\slashed{p}_{11}+m_{Q}\right).
\end{eqnarray}

The $P$-wave amplitudes and spin projector can be obtained by the first derivation of the relative momentum $p$ in $S$-wave amplitudes, i.e., 
\begin{eqnarray}
		M_{n}[^1 P_1]&= & \varepsilon_\alpha^l(p_1) \frac{d}{d p_\alpha}
        M_{n}[^1 S_0]|_{p=0},\\
		M_{n}[^3 P_J]&= & \varepsilon_{\alpha \beta}^J(p_1) \frac{d}{d p_\alpha}M_{n}^{\beta}[^3 S_1]|_{p=0},\\
\left. \frac{d}{dp_{\alpha}}\Pi[^1 S_0](p_1) \right|_{p=0}&=& \frac{\sqrt{m_{QQ^{\prime}}}}{4 m_{Q} m_{Q^{\prime}}}\gamma^{\alpha} \gamma^{5} \left(\slashed{p}_{1}+m_{Q}-m_{Q^{\prime}}\right), \\
\left. \frac{d}{dp_{\alpha}}\Pi[^3 S_1]^{\beta}(p_1) \right|_{p=0}&=& \frac{\sqrt{m_{QQ^{\prime}}}}{4 m_{Q} m_{Q^{\prime}}} \left[ \gamma^{\alpha} \gamma^{\beta} \left(\slashed{p}_{1}+m_{Q}-m_{Q^{\prime}}\right) -2 g^{\alpha \beta} \left(\slashed{p}_{12} - m_{Q^{\prime}} \right) \right].
\end{eqnarray}

In the amplitude $M_{n}[^1 S_0]$ and $M_{n}^{\beta}[^3 S_1]$, the relative momentum $p$, which is contained in the constituent quark momenta of the diquark, $p_{11}$ and $p_{12}$, always appears in the projector and propagators. $\varepsilon_\beta^s(p_1)$ and $\varepsilon_\alpha^l(p_1)$ correspond to the polarization vectors of the spin and orbital angular momentum of the diquark in $[^3S_1]$- and $[^1 P_1]$-state respectively. $\varepsilon_{\alpha \beta}^J(p_1)$ is the polarization tensor of $[^3 P_J]$- states with $J=0, 1,$ or $2$. All of them need to be polarization summed to select the suitable total angular momentum, and the polarization summation formulas of polarization vector and tensor satisfy \cite{Petrelli:1997ge}

\begin{eqnarray}
	&&\sum_{l_z} \varepsilon_\alpha^l \varepsilon_{\alpha^{\prime}}^{l *}=\Pi_{\alpha \alpha^{\prime}},\\
	&&\varepsilon_{\alpha \beta}^0 \varepsilon_{\alpha^{\prime} \beta^{\prime}}^{0 *}=\frac{1}{3} \Pi_{\alpha \beta} \Pi_{\alpha^{\prime} \beta^{\prime}},\\
	&&\sum_{J_z} \varepsilon_{\alpha \beta}^1 \varepsilon_{\alpha^{\prime} \beta^{\prime}}^{1 *}=\frac{1}{2}(\Pi_{\alpha \alpha^{\prime}} \Pi_{\beta \beta^{\prime}}-\Pi_{\alpha \beta^{\prime}} \Pi_{\alpha^{\prime} \beta}), \\
	&&\sum_{J_z} \varepsilon_{\alpha \beta}^2 \varepsilon_{\alpha^{\prime} \beta^{\prime}}^{2 *}=\frac{1}{2}(\Pi_{\alpha \alpha^{\prime}} \Pi_{\beta \beta^{\prime}}+\Pi_{\alpha \beta^{\prime}} \Pi_{\alpha^{\prime} \beta})-\frac{1}{3} \Pi_{\alpha \beta} \Pi_{\alpha^{\prime} \beta^{\prime}} ,
\end{eqnarray}
with the definition $\Pi_{\alpha \beta}=-g_{\alpha \beta}+\frac{p_{1 \alpha} p_{1 \beta}}{m_{QQ^{\prime}}^2}$.

The color factor $\mathcal{C}_{i j, k}$ of the diquark has been extracted from the amplitudes and it satisfies
\begin{equation}
\mathcal{C}_{i j, k}=\mathcal{N} \times \sum_{a=1}^8 \sum_{m, n=1}^3 (T^a)_{m i}(T^a)_{n j} \times G_{m n k},
\end{equation}
in which $\mathcal{N}=1/\sqrt{2}$ is the normalization constant, $i,j$ and $m,n$ are the color indices of two heavy antiquark and quarks, $k$ denotes the color index of the diquark, and $G_{m n k}$ can be considered
equal to the antisymmetric function $\varepsilon_{m n k}$ for $\bar{\mathbf 3}$ state, or the symmetric function $f_{m n k}$ for $\mathbf 6$ state, which can be
\begin{eqnarray}
	\varepsilon_{m n k} \varepsilon_{m^{\prime} n^{\prime} k}=\delta_{m m^{\prime}} \delta_{n n^{\prime}}-\delta_{m n^{\prime}} \delta_{n m^{\prime}},\\
	f_{m n k} f_{m^{\prime} n^{\prime} k}=\delta_{m m^{\prime}} \delta_{n n^{\prime}}+\delta_{m n^{\prime}} \delta_{n m^{\prime}}.
\end{eqnarray}
Thus, the color factors are numerically calculated to be $4/3$ for color-$\bar{\textbf{3}}$ state, and $2/3$ for $\textbf{6}$ state.

\subsection{Long-distance matrix elements}

The hadronic process represented by the nonperturbative long-distance matrix elements $\langle{\cal O}^{H}[n]\rangle$ can be divided into two steps: first the perturbative diquark pair $\langle QQ^{\prime} \rangle[n]$ is bound to the diquark state, which can be described by the transition probability $h_{\bar{\textbf{3}}}$ and $h_{\textbf{6}}$ for color-antitriplet and sextuplet diquark respectively, and then the diquark state captures a light quark ($q=u, d$,~or, $s$) from the vacuum to form a baryon $\Xi_{QQ^{\prime}}$ with  $100\%$ efficiency. Here the isospin-breaking effect is
ignored, and for simplicity, we use $\Xi_{QQ^{\prime}}$ to denote $\Xi_{QQ^{\prime}q}$. Typically assuming the potential of the binding color-antitriplet diquark state is hydrogenlike, $h_{\bar{\textbf{3}}}$ can be approximately associated with the Schr\"{o}dinger wave function at the origin $|\Psi_{QQ^{\prime}}(0)|^2$ for $S$-wave diquark states, and the first-derivative wave function at the origin $|\Psi_{QQ^{\prime}}^{\prime}(0)|^2$ for $P$-wave states. And the (first-derivative) wave function at the origin
can naturally connect with the (derivative) radial wave
function at the origin, i.e.,
\begin{eqnarray}
h[S]_{\bar{\textbf{3}}} \simeq |\Psi_{QQ^{\prime}}(0)|^2 &=&\frac{1}{4\pi}|R_{QQ^{\prime}}(0)|^2, \nonumber\\
h[P]_{\bar{\textbf{3}}}\simeq|\Psi^{\prime}_{QQ^{\prime}}(0)|^2&=&\frac{3}{4\pi}|R^{\prime}_{QQ^{\prime}}(0)|^2.
\label{eq:h3}
\end{eqnarray}

The numerical value of $|R_{QQ^{\prime}}(0)|^2$ and $|R^{\prime}_{QQ^{\prime}}(0)|^2$ can be estimated by fitting the experimental data or some nonperturbative methods like QCD sum rules~\cite{Kiselev:1999sc}, lattice QCD~\cite{Bodwin:1996tg} or the potential model~\cite{Bagan:1994dy}. Its values are slightly different obtained in different potential model, such as Power-law~\cite{Bagan:1994dy}, $\rm K^2O$ potential~\cite{Kiselev:2002iy}, and $\rm Buchm\ddot{u}ller$-$\rm Tye~(B.T.)$~\cite{Kiselev:2001fw}. Fortunately, at the present considered pQCD level, the wave function at the origin and its first-derivation can be regarded as a global factor to update the numerical results if more accurate values of the wave function at the origin are available.

As for $h_{\textbf{6}}$, according to NRQCD's power counting rules, there are two different arguments compared to $h_{\bar{\textbf{3}}}$. 
In NRQCD theory, $\Xi_{QQ^{\prime}}$ can be expanded into a series of Fock states with a small relative velocity $v$ between heavy quarks in the rest frame of the diquark:
\begin{equation}
	\label{eq:fock}
	\begin{aligned}
		\left|\Xi_{QQ^{\prime}}\right\rangle= c_1(v)|(QQ^{\prime}) q\rangle+c_2(v)|(QQ^{\prime}) q g\rangle 
		 +c_3(v)|(QQ^{\prime}) q g g\rangle+\cdots.
	\end{aligned}
\end{equation}

Consider a diquark $\langle QQ^{\prime}\rangle_{\bar{\textbf{3}}}$ to form a baryon $|(QQ^{\prime}) q\rangle$, one heavy quark of diquark can emit a gluon without changing the spin of the heavy quark. Subsequently, this emitted gluon can split into a light quark-antiquark pair $q\bar{q}$. The resulting light $q$ interact with the diquark $\langle QQ^{\prime}\rangle_{\bar{\textbf{3}}}$ then to form the baryon $\Xi_{QQ^{\prime}}$. However, during the formation of baryon $|(QQ^{\prime}) q\rangle$ from the diquark $\langle QQ^{\prime}\rangle_{\textbf{6}}$, the process is similar to $\langle QQ^{\prime}\rangle_{\bar{\textbf{3}}}$, except that the emitted gluon must change the spin of the heavy quark, leading to a suppression to $h_{\textbf{6}}$ \cite{Zheng:2015ixa}. 

Another assumption is that the diquark $\langle QQ^{\prime}\rangle_{\textbf{6}}$ can also form baryon $|(QQ^{\prime}) qg\rangle$. The process is that one heavy quarks in diquark emits a gluon without changing the spin of the heavy quark, and then the light antiquark formed by the emitted gluon splitting emit a gluon again. Finally the gluon and light quark are captured by the diquark to form $|(QQ^{\prime}) qg\rangle$. This makes the contribution of $h_{\textbf{6}}$ equal to $h_{\bar{\textbf{3}}}$. Therefore, since a light quark can easily emit gluons, the constituents in Eq.~(\ref{eq:fock}) are equivalent significance, i.e., $c_1\sim c_2 \sim c_3$ \cite{Ma:2003zk}. So we adopt  $h_{\textbf{6}}=h_{\bar{\textbf{3}}}$ in the subsequent numerical calculation.

\section{ Numerical results}\label{sec3}
The input parameters in the numerical calculation are listed below \cite{Baranov:1995rc,ParticleDataGroup:2020ssz}:
\begin{eqnarray}
&&m_c=1.8~\rm{GeV},~~~~~~~~~~~\it{m_b}=\rm 5.1~{GeV},~~~~~~~~~~~~\it{m_{QQ^{\prime}}}=m_Q+m_{Q^{\prime}},\nonumber\\
&&m_Z=91.1876~\rm{GeV},~~~~~\it{m_W}=\rm 80.385~{GeV},~~~~~~G_{F}=1.1663787 \times 10^{-5},\nonumber\\
&&|R_{cc}(0)|=0.700~{\rm GeV}^{\frac{3}{2}},~|R_{bc}(0)|=0.904~{\rm GeV}^{\frac{3}{2}},~|R_{bb}(0)|=1.382~{\rm GeV}^{\frac{3}{2}},\nonumber\\
&&|R_{cc}^{\prime}(0)|=0.102~{\rm GeV}^{\frac{5}{2}},~|R_{bc}^{\prime}(0)|=0.200~{\rm GeV}^{\frac{5}{2}},~|R_{bb}^{\prime}(0)|=0.479~{\rm GeV}^{\frac{5}{2}},
\end{eqnarray}
where $|R_{QQ^{\prime}}^{\prime}(0)|$ are evaluated under the $\rm K^2O$ potential motivated by QCD with a three-loop function \cite{Kiselev:2002iy}. The renormalization scale is typically taken as the transverse mass of the final-state doubly heavy baryon, specifically $\mu = \sqrt{M^2_{\Xi_{QQ^{\prime}}} + p^2_T}$, where $p_T$ denotes the transverse momentum of the particle. Accordingly, the strong running coupling can be obtained from the solution of the one-loop renormalization group equation with the
reference point $\alpha_s(m_Z)$=0.1181 \cite{Baikov:2016tgj,Herzog:2017ohr}. The programs FeynArts 3.9 \cite{Hahn:2000kx}, FeynCalc 9.3 \cite{Shtabovenko:2020gxv}, and the modified FormCalc \cite{Hahn:1998yk} are
used to generate the amplitudes and do the algebraic and
numerical calculations.

\subsection{Cross section}

\begin{figure}
\centering
    \includegraphics[scale=0.5]{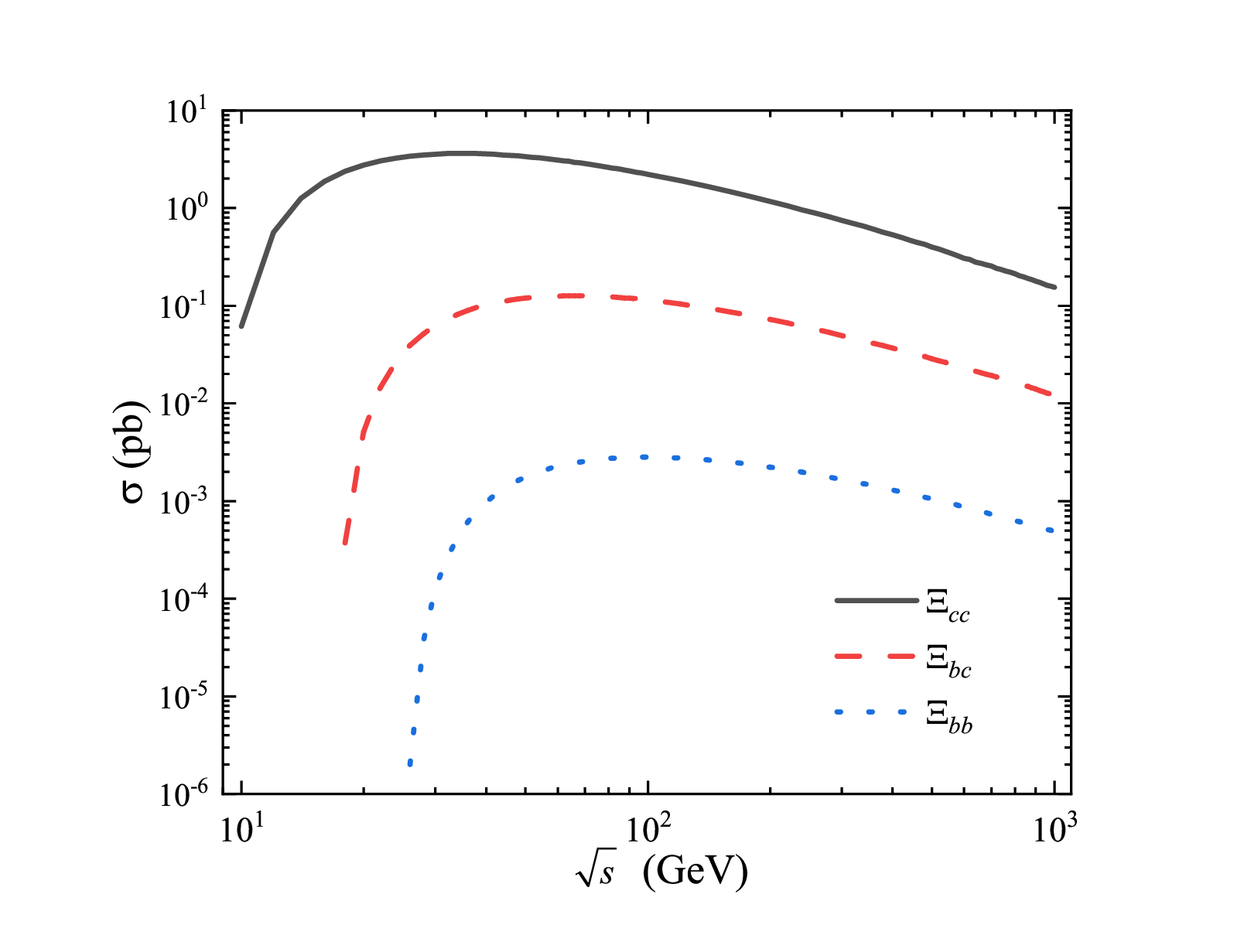}
\caption{Energy dependence of the cross section for the  photoproduction of $\Xi_{QQ^{\prime}}$ in $S$-wave.} 
\label{energyd}
\end{figure}

The energy dependence of the cross section for the $S$-wave $\Xi_{QQ^{\prime}}$ photoproduction is shown in Fig.~\ref{energyd}. It can be intuitively seen from Fig.~\ref{energyd} that with the increase of the collision energy $\sqrt{s}$, the cross section for the photoproduction of $\Xi_{QQ^{\prime}}$ initially increases rapidly and subsequently decreases, and the maximum value appears in the energy range of tens of GeV.

The cross sections for the photoproduction of $\Xi_{cc}$, $\Xi_{bc}$, and $\Xi_{bb}$ from different configurations of intermediate diquark have been analyzed under four typical collision energies $\sqrt{s}=91,~160,~240$, and 360 GeV, and the corresponding results are given in Tables~\ref{mcc}-\ref{mbb}. Assuming that the highly excited states transition $100\%$ to the ground state, the total cross section of $\Xi_{QQ^{\prime}}$ photoproduction can be obtained by summing up the contributions of the intermediate diquark configurations. 
Tables~\ref{mcc}-\ref{mbb} show that 

\begin{itemize}
    \item The total cross sections decrease with the increment of $\sqrt{s}$, which is consistent with the trend in Fig.~\ref{energyd}.
    \item The ratio of the cross sections obtained by four different collision energy $\sqrt{s}$ (unit: GeV) is $\sigma(91):\sigma(160):\sigma(240):\sigma(360)\simeq$  $3.92:2.40:1.59:1$, $2.87:2.08:1.49:1$, and $1.96:1.75:1.38:1$ for the photoproduction of $\Xi_{cc}$, $\Xi_{bc}$, and $\Xi_{bb}$, respectively.
    \item The biggest contribution comes from the intermediate states with quantum number $[^3S_1]_{\mathbf{\bar 3}}$.
    \item Comparing the contributions between $P$-wave states, the largest result occurs in $[^1P_1]_{\mathbf{\bar 3}}$ for the photoproduction of $\Xi_{cc}$ and $\Xi_{bb}$, $[^3P_2]_{\bar{\mathbf 3}}$ for $\Xi_{bc}$.
    \item At $\sqrt{s}=91$~GeV, the contribution from $P$-wave states is approximately $2.19\%$, $4.23\%$, $1.26\%$ of the $S$-wave contribution for the photoproduction of $\Xi_{cc}$, $\Xi_{bc}$ and $\Xi_{bb}$, respectively. As the collision energy increases, the contribution of $P$-wave also increases, up to $2.46\%$, $4.96\%$, $1.60\%$ when $\sqrt{s}=$ 360 GeV.
\end{itemize}

\begin{table}[htb]
\begin{center}
\caption{Cross sections (in unit: pb) for the photoproduction of $\Xi_{cc}$ under four typical collision energies $\sqrt{s}=91,~160,~240$, and 360 GeV.} \vspace{0.5cm}
\begin{tabular}{|cc|c|c|c|c|}
\hline
\multicolumn{2}{|c|}{$\sqrt{s}$ (GeV)}                                      & 91                     & 160                     & 240           & 360          \\ \hline
\multicolumn{1}{|c|}{\multirow{3}{*}{$S$-wave}} & $[^1S_0]_{\mathbf 6}$               & $1.58 ~\times ~10^{-1}$ & $1.12 ~\times ~10^{-1}$ & $7.90 ~\times ~10^{-2}$ & $5.29~\times ~10^{-2}$   \\ \cline{2-6} 
\multicolumn{1}{|c|}{}                        & $[^3S_1]_{\mathbf{\bar 3}}$ &         $2.24 $         & $1.36$             
          & $8.90 ~\times ~10^{-1}$ & $5.57 ~\times ~10^{-1}$  \\ \cline{2-6} 
\multicolumn{1}{|c|}{}                        & $S$-total                     &          $2.40$         & $1.47$            
          & $9.69~\times ~10^{-1}$  & $6.10 ~\times ~10^{-1}$\\ \hline
\multicolumn{1}{|c|}{\multirow{5}{*}{$P$-wave}} & $[^1P_1]_{\mathbf{\bar 3}}$ & $2.70 ~\times ~10^{-2}$ & $1.70 ~\times ~10^{-2}$ & $1.13 ~\times ~10^{-2}$ & $7.17 ~\times ~10^{-3}$ \\ \cline{2-6} 
\multicolumn{1}{|c|}{}                        & $[^3P_0]_{\mathbf 6}$               & $3.99 ~\times ~10^{-3}$ & $2.55 ~\times ~10^{-3}$ & $1.72 ~\times ~10^{-3}$ & $1.10 ~\times ~10^{-3}$ \\ \cline{2-6} 
\multicolumn{1}{|c|}{}                        & $[^3P_1]_{\mathbf 6}$               & $4.89 ~\times ~10^{-3}$ & $3.42 ~\times ~10^{-3}$ & $2.41 ~\times ~10^{-3}$ & $1.60 ~\times ~10^{-3}$ \\ \cline{2-6} 
\multicolumn{1}{|c|}{}                        & $[^3P_2]_{\mathbf 6}$               & $1.66 ~\times ~10^{-2}$ & $1.12 ~\times ~10^{-2}$ & $7.77 ~\times ~10^{-3}$ & $5.10 ~\times ~10^{-3}$\\ \cline{2-6} 
\multicolumn{1}{|c|}{}                        & $P$-total                     & $5.25 ~\times ~10^{-2}$ & $3.42 ~\times ~10^{-2}$ & $2.32 ~\times ~10^{-2}$  & $1.50 ~\times ~10^{-2}$\\ \hline
\multicolumn{2}{|c|}{Total}                   &           $2.45$            &          $1.50$         & $9.92 ~\times ~10^{-1}$ & $6.25 ~\times ~10^{-1}$ \\ \hline
\end{tabular}
\label{mcc}
\end{center}
\end{table}

\begin{table}[htb]
\begin{center}
\caption{Cross sections (in unit: pb) for the photoproduction of $\Xi_{bc}$ under four typical collision energies $\sqrt{s}=91,~160,~240$, and 360 GeV.} \vspace{0.5cm}
\begin{tabular}{|cc|c|c|c|c|}
\hline
\multicolumn{2}{|c|}{$\sqrt{s}$ (GeV)}                                      & 91                     & 160                     & 240                     & 360 \\ \hline
\multicolumn{1}{|c|}{\multirow{5}{*}{$S$-wave}} & $[^1S_0]_{\mathbf 6}$               & $1.23 ~\times ~10^{-2}$ & $9.03 ~\times ~10^{-3}$ & $6.43 ~\times ~10^{-3}$ & $4.29 ~\times ~10^{-3}$ \\ \cline{2-6} 
\multicolumn{1}{|c|}{}                        & $[^1S_0]_{\mathbf{\bar 3}}$ & $2.47 ~\times ~10^{-2}$ & $1.81 ~\times ~10^{-2}$ & $1.29 ~\times ~10^{-2}$ & $8.58 ~\times ~10^{-3}$ \\ \cline{2-6} 
\multicolumn{1}{|c|}{}                        & $[^3S_1]_{\mathbf 6}$               & $2.73 ~\times ~10^{-2}$ & $1.98 ~\times ~10^{-2}$ & $1.42 ~\times ~10^{-2}$ & $9.57 ~\times ~10^{-3}$ \\ \cline{2-6} 
\multicolumn{1}{|c|}{}                        & $[^3S_1]_{\mathbf{\bar 3}}$ & $5.47 ~\times ~10^{-2}$ & $3.97 ~\times ~10^{-2}$ & $2.84 ~\times ~10^{-2}$ & $1.91 ~\times ~10^{-2}$ \\ \cline{2-6} 
\multicolumn{1}{|c|}{}                        & $S$-total & $1.20 ~\times ~10^{-1}$ & $8.66 ~\times ~10^{-2}$ & $6.19 ~\times ~10^{-2}$ & $4.15 ~\times ~10^{-2}$ \\ \hline
\multicolumn{1}{|c|}{\multirow{9}{*}{$P$-wave}} & $[^1P_1]_{\mathbf 6}$               & $3.39 ~\times ~10^{-4}$ & $2.74 ~\times ~10^{-4}$ & $2.07 ~\times ~10^{-4}$ & $1.44 ~\times ~10^{-4}$ \\ \cline{2-6} 
\multicolumn{1}{|c|}{}                        & $[^1P_1]_{\mathbf{\bar 3}}$ & $6.78 ~\times ~10^{-4}$ & $5.48 ~\times ~10^{-4}$ & $4.13 ~\times ~10^{-4}$ & $2.88 ~\times ~10^{-4}$\\ \cline{2-6} 
\multicolumn{1}{|c|}{}                        & $[^3P_0]_{\mathbf 6}$               & $1.02 ~\times ~10^{-4}$ & $7.81 ~\times ~10^{-5}$ & $5.74 ~\times ~10^{-5}$ & $3.94 ~\times ~10^{-5}$ \\ \cline{2-6} 
\multicolumn{1}{|c|}{}                        & $[^3P_0]_{\mathbf{\bar 3}}$ & $2.04 ~\times ~10^{-4}$ & $1.56 ~\times ~10^{-4}$ & $1.15 ~\times ~10^{-4}$ & $7.88 ~\times ~10^{-5}$ \\ \cline{2-6} 
\multicolumn{1}{|c|}{}                        & $[^3P_1]_{\mathbf 6}$               & $2.97 ~\times ~10^{-4}$ & $2.32 ~\times ~10^{-4}$ & $1.70 ~\times ~10^{-4}$ & $1.16 ~\times ~10^{-4}$\\ \cline{2-6} 
\multicolumn{1}{|c|}{}                        & $[^3P_1]_{\mathbf{\bar 3}}$ & $5.95 ~\times ~10^{-4}$ & $4.63 ~\times ~10^{-4}$ & $3.40 ~\times ~10^{-4}$ & $2.33 ~\times ~10^{-4}$\\ \cline{2-6} 
\multicolumn{1}{|c|}{}                        & $[^3P_2]_{\mathbf 6}$               & $9.49 ~\times ~10^{-4}$ & $7.58 ~\times ~10^{-4}$ & $5.63 ~\times ~10^{-4}$ & $3.87 ~\times ~10^{-4}$ \\ \cline{2-6} 
\multicolumn{1}{|c|}{}                        & $[^3P_2]_{\mathbf{\bar 3}}$ & $1.90 ~\times ~10^{-3}$ & $1.52 ~\times ~10^{-3}$ & $1.13 ~\times ~10^{-3}$ & $7.74 ~\times ~10^{-4}$ \\ \cline{2-6}
\multicolumn{1}{|c|}{}                        & $P$-total & $5.07 ~\times ~10^{-3}$ & $4.03 ~\times ~10^{-3}$ & $3.00 ~\times ~10^{-3}$ & $2.06 ~\times ~10^{-3}$ \\ \hline
\multicolumn{2}{|c|}{Total}     & $1.25 ~\times ~10^{-1}$     & $9.06 ~\times ~10^{-2}$   & $6.49 ~\times ~10^{-2}$                     &  $4.36 ~\times ~10^{-2}$ \\ \hline
\end{tabular}
\label{mbc}
\end{center}
\end{table}

\begin{table}[htb]
\begin{center}
\caption{Cross sections (in unit: pb) for the photoproduction of $\Xi_{bb}$ under four typical collision energies $\sqrt{s}=91,~160,~240$, and 360 GeV.} \vspace{0.5cm}
\begin{tabular}{|cc|c|c|c|c|}
\hline
\multicolumn{2}{|c|}{$\sqrt{s}$ (GeV)}                                      & 91                     & 160                     & 240           & 360          \\ \hline
\multicolumn{1}{|c|}{\multirow{3}{*}{$S$-wave}} & $[^1S_0]_{\mathbf 6}$               & $1.17 ~\times ~10^{-4}$ & $1.41 ~\times ~10^{-4}$ & $1.29 ~\times ~10^{-4}$ & $1.04 ~\times ~10^{-4}$   \\ \cline{2-6} 
\multicolumn{1}{|c|}{}                        & $[^3S_1]_{\mathbf{\bar 3}}$ & $2.68 ~\times ~10^{-3}$ & $2.36 ~\times ~10^{-3}$ & $1.84 ~\times ~10^{-3}$ & $1.33 ~\times ~10^{-3}$  \\ \cline{2-6} 
\multicolumn{1}{|c|}{}                        & $S$-total                     & $2.80 ~\times ~10^{-3}$ & $2.50 ~\times ~10^{-3}$ & $1.97 ~\times ~10^{-3}$ & $1.43 ~\times ~10^{-3}$\\ \hline
\multicolumn{1}{|c|}{\multirow{5}{*}{$P$-wave}} & $[^1P_1]_{\mathbf{\bar 3}}$ & $2.05 ~\times ~10^{-5}$ & $1.94 ~\times ~10^{-5}$ & $1.57 ~\times ~10^{-5}$ & $1.16 ~\times ~10^{-5}$ \\ \cline{2-6} 
\multicolumn{1}{|c|}{}                        & $[^3P_0]_{\mathbf 6}$               & $2.79 ~\times ~10^{-6}$ & $2.79 ~\times ~10^{-6}$ & $2.31 ~\times ~10^{-6}$ & $1.73 ~\times ~10^{-6}$ \\ \cline{2-6} 
\multicolumn{1}{|c|}{}                        & $[^3P_1]_{\mathbf 6}$               & $2.53 ~\times ~10^{-6}$ & $3.10 ~\times ~10^{-6}$ & $2.82 ~\times ~10^{-6}$ & $2.27 ~\times ~10^{-6}$ \\ \cline{2-6} 
\multicolumn{1}{|c|}{}                        & $[^3P_2]_{\mathbf 6}$               & $9.36 ~\times ~10^{-6}$ & $1.06 ~\times ~10^{-5}$ & $9.35 ~\times ~10^{-6}$ & $7.32 ~\times ~10^{-6}$\\ \cline{2-6} 
\multicolumn{1}{|c|}{}                        & $P$-total                     & $3.52 ~\times ~10^{-5}$ & $3.59 ~\times ~10^{-5}$ & $3.02 ~\times ~10^{-5}$  & $2.29 ~\times ~10^{-5}$\\ \hline
\multicolumn{2}{|c|}{Total}     &  $2.84 ~\times ~10^{-3}$    &  $2.54 ~\times ~10^{-3}$  &  $2.00 ~\times ~10^{-3}$ & $1.45 ~\times ~10^{-3}$ \\ \hline
\end{tabular}
\label{mbb}
\end{center}
\end{table}

Using the total cross section calculated above, we can estimate the events of doubly heavy baryons that can be produced at the CEPC and FCC-ee by formula $N=\sigma\times\mathcal{L}$. The designed luminosity $\mathcal{L}$ of CEPC and FCC-ee both with two interaction points (2IPs) at different collision energies \cite{CEPCStudyGroup:2023quu,FCC:2018evy} is shown in Table \ref{Lumi}. Based on these collision conditions (luminosity and collision energy), the predicted $\Xi_{QQ^{\prime}}$ events produced at CEPC and FCC-ee per year are calculated in Table~\ref{events}.


\begin{table}[htb]
\begin{center}
\caption{Designed Luminosity $\mathcal{L}$ of CEPC and FCC-ee at different collision energies.} \vspace{0.5cm}
\begin{tabular}{|c|ccc|}
\hline
\multirow{2}{*}{$\sqrt{s}$ (GeV)} & \multicolumn{3}{c|}{$\mathcal{L}$~$\times 10^{34}~cm^{-2}s^{-1}$}                                                \\ \cline{2-4} 
                  & \multicolumn{1}{c|}{CEPC@30MW} & \multicolumn{1}{c|}{CEPC@50MW} & FCC-ee \\ \hline
91                & \multicolumn{1}{c|}{115}       & \multicolumn{1}{c|}{192}       & 460    \\ \hline
160               & \multicolumn{1}{c|}{16}        & \multicolumn{1}{c|}{27}        & 56     \\ \hline
240               & \multicolumn{1}{c|}{5}         & \multicolumn{1}{c|}{8.3}       & 17     \\ \hline
360               & \multicolumn{1}{c|}{0.5}       & \multicolumn{1}{c|}{0.8}       & 3.1    \\ \hline
\end{tabular}
\label{Lumi}
\end{center}
\end{table}

\begin{table}[htb]
\begin{center}
\caption{Predicted $\Xi_{QQ^{\prime}}$ events produced at CEPC and FCC-ee per year based on the collision conditions.} \vspace{0.5cm}
\begin{tabular}{|cc|c|c|c|c|}
\hline
\multicolumn{2}{|c|}{\diagbox{Events}{$\sqrt{S}$ (GeV)}} & 91 & 160 & 240 & 360 \\ \hline
\multicolumn{1}{|c|}{\multirow{3}{*}{$\Xi_{cc}$}} & CEPC@30MW &  $8.90 \times 10^7$  & $7.59 \times 10^6$    &  $1.56 \times 10^6$   & $9.86 \times 10^4$    \\ \cline{2-6} 
\multicolumn{1}{|c|}{}                      & CEPC@50MW & $1.49 \times 10^8$   &  $1.28 \times 10^7$   & $2.60 \times 10^6$    &  $1.58 \times 10^5$   \\ \cline{2-6} 
\multicolumn{1}{|c|}{}                      & FCC-ee    & $3.56 \times 10^8$   &  $2.66 \times 10^7$   &  $5.32 \times 10^6$   &  $6.11 \times 10^5$   \\ \hline
\multicolumn{1}{|c|}{\multirow{3}{*}{$\Xi_{bc}$}} & CEPC@30MW &  $4.54 \times 10^6$  &  $4.57 \times 10^5$   & $1.02 \times 10^5$   &  $6.87 \times 10^3$   \\ \cline{2-6} 
\multicolumn{1}{|c|}{}                      & CEPC@50MW & $7.57 \times 10^6$   &  $7.72 \times 10^5$   & $1.70 \times 10^5$    &   $1.10 \times 10^4$    \\ \cline{2-6} 
\multicolumn{1}{|c|}{}                      & FCC-ee    & $1.81 \times 10^7$   &  $1.60 \times 10^6$   &  $3.48 \times 10^5$   &  $4.26 \times 10^4$   \\ \hline
\multicolumn{1}{|c|}{\multirow{3}{*}{$\Xi_{bb}$}} & CEPC@30MW &  $1.03 \times 10^5$  &  $1.28 \times 10^4$   & $3.15 \times 10^3$     & $2.29 \times 10^2$    \\ \cline{2-6} 
\multicolumn{1}{|c|}{}                      & CEPC@50MW & $1.72 \times 10^5$   &  $2.16 \times 10^4$    &  $5.24 \times 10^3$   & $3.67 \times 10^2$    \\ \cline{2-6} 
\multicolumn{1}{|c|}{}                      & FCC-ee    & $4.11 \times 10^5$   &  $4.48 \times 10^4$    &  $1.07 \times 10^4$   &  $1.42 \times 10^3$   \\ \hline
\end{tabular}
\label{events}
\end{center}
\end{table}

From Table~\ref{events}, we can see that both at CEPC and FCC-ee, a large number of $\Xi_{QQ^{\prime}}$ events can be produced. In particular, when $\sqrt{s} = 91$~GeV, the number of produced $\Xi_{cc}$, $\Xi_{bc}$, and $\Xi_{bb}$ events is as high as $\mathcal{O}(10^8)$, $\mathcal{O}(10^7),$ and $\mathcal{O}(10^5)$ respectively, which is very promising to be detected in future experiments. At the same luminosity, the ratios of $\Xi_{cc}$, $\Xi_{bc}$, and $\Xi_{bb}$ events produced at the collision energies $\sqrt{s} = 91, 160, 240, 360$ are about $863:44:1$, $591:36:1$, $496:32:1$, and $431:30:1$, respectively.
The produced events of $\Xi_{bc}$ ($\Xi_{bb}$) is one (two) order(s) of magnitude smaller than that of $\Xi_{cc}$. 

\subsection{Differential distributions}

\begin{figure}
\centering
\hspace{-0.50in}
    \includegraphics[scale=0.22]{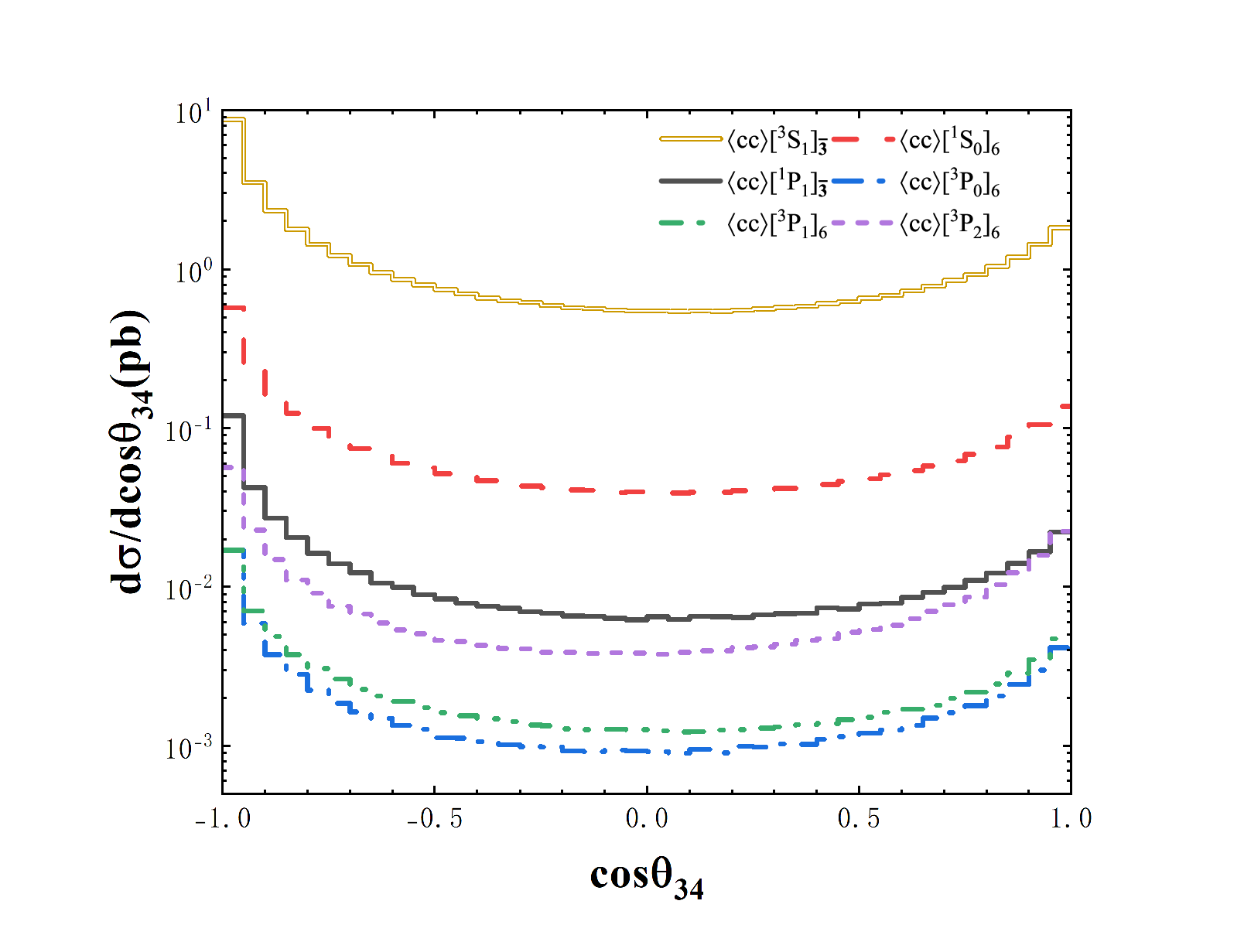}
  \hspace{-0.50in}
    \includegraphics[scale=0.22]{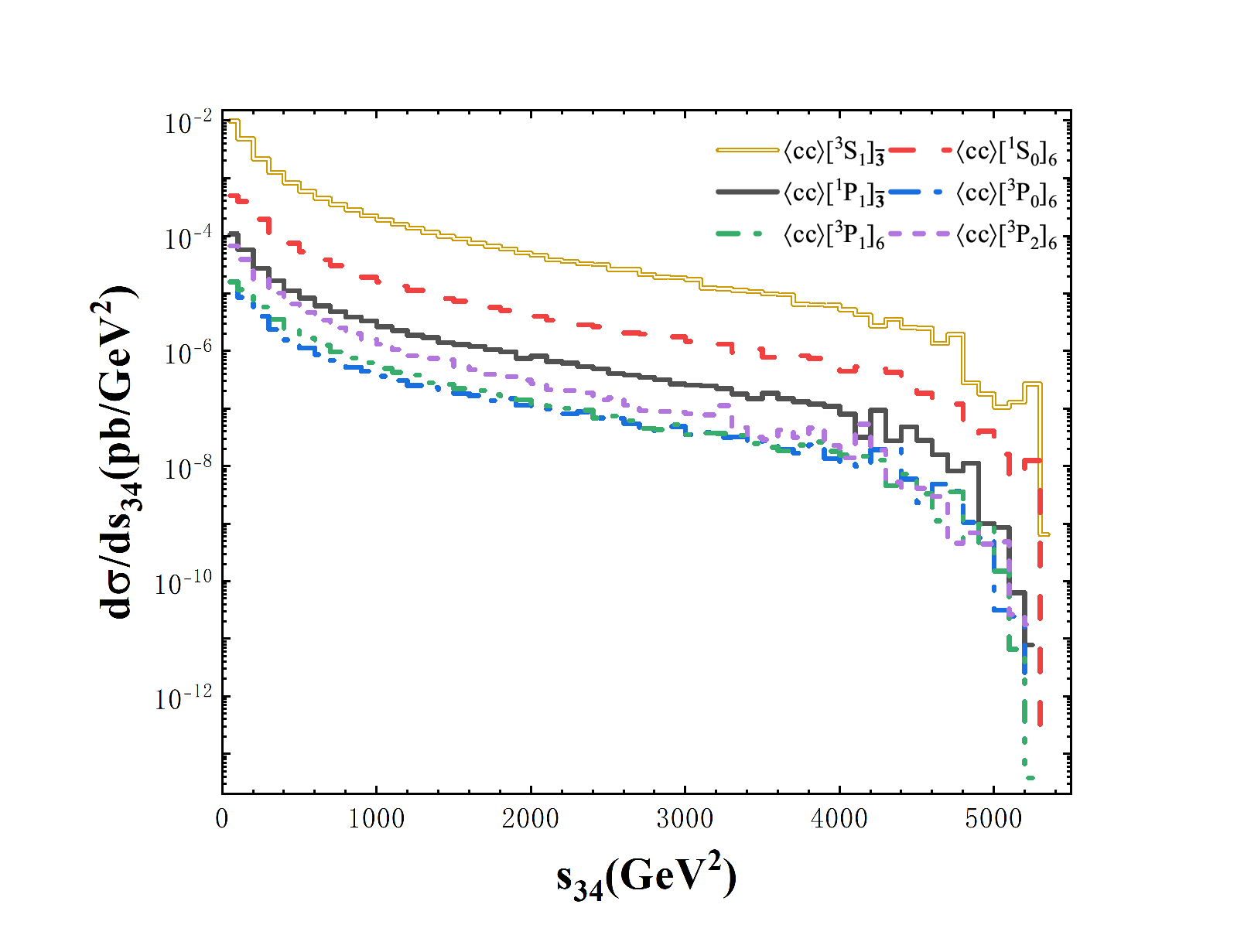}
  \hspace{-0.50in}
    \includegraphics[scale=0.22]{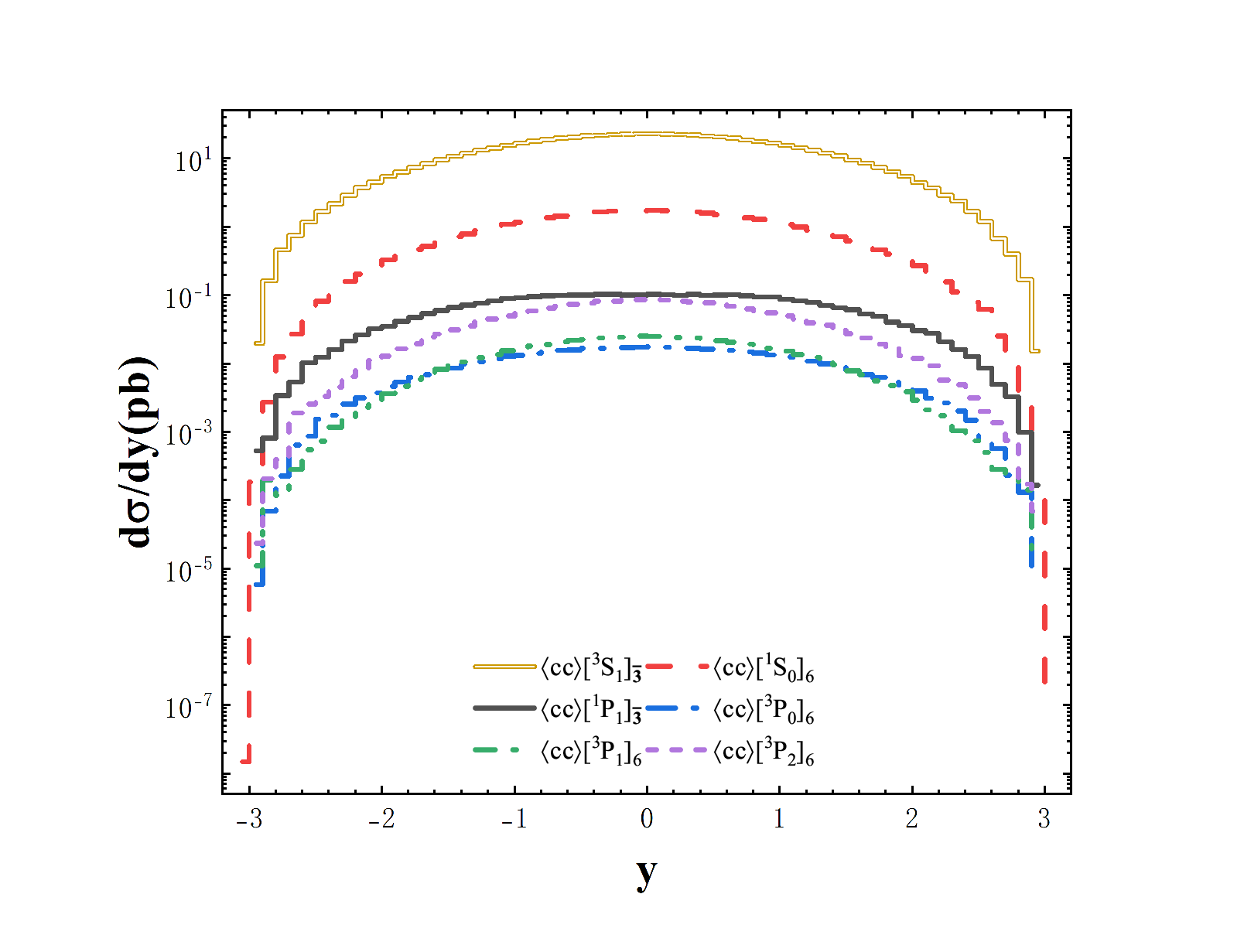}\\
\hspace{-0.50in}
    \includegraphics[scale=0.22]{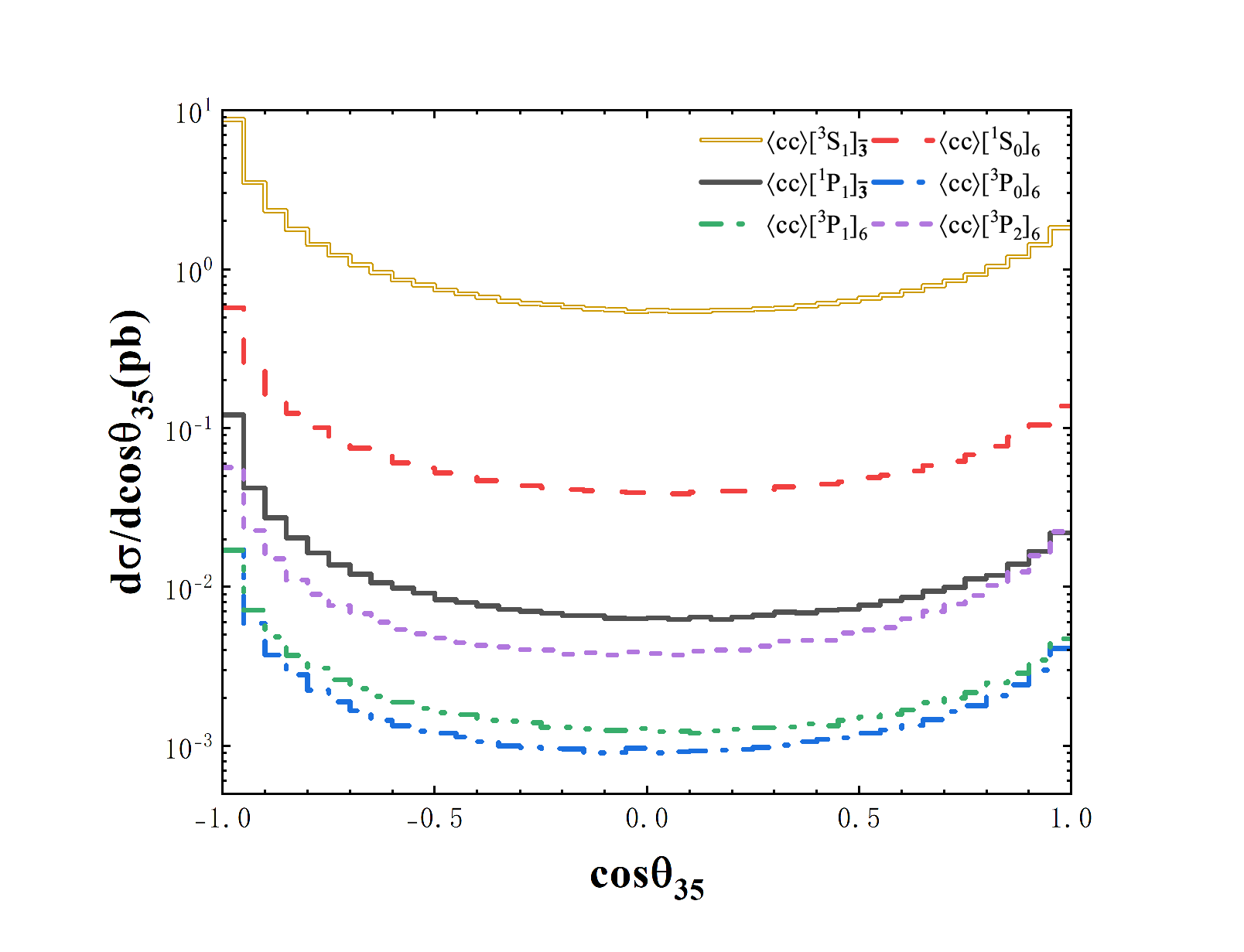}
\hspace{-0.50in}
    \includegraphics[scale=0.22]{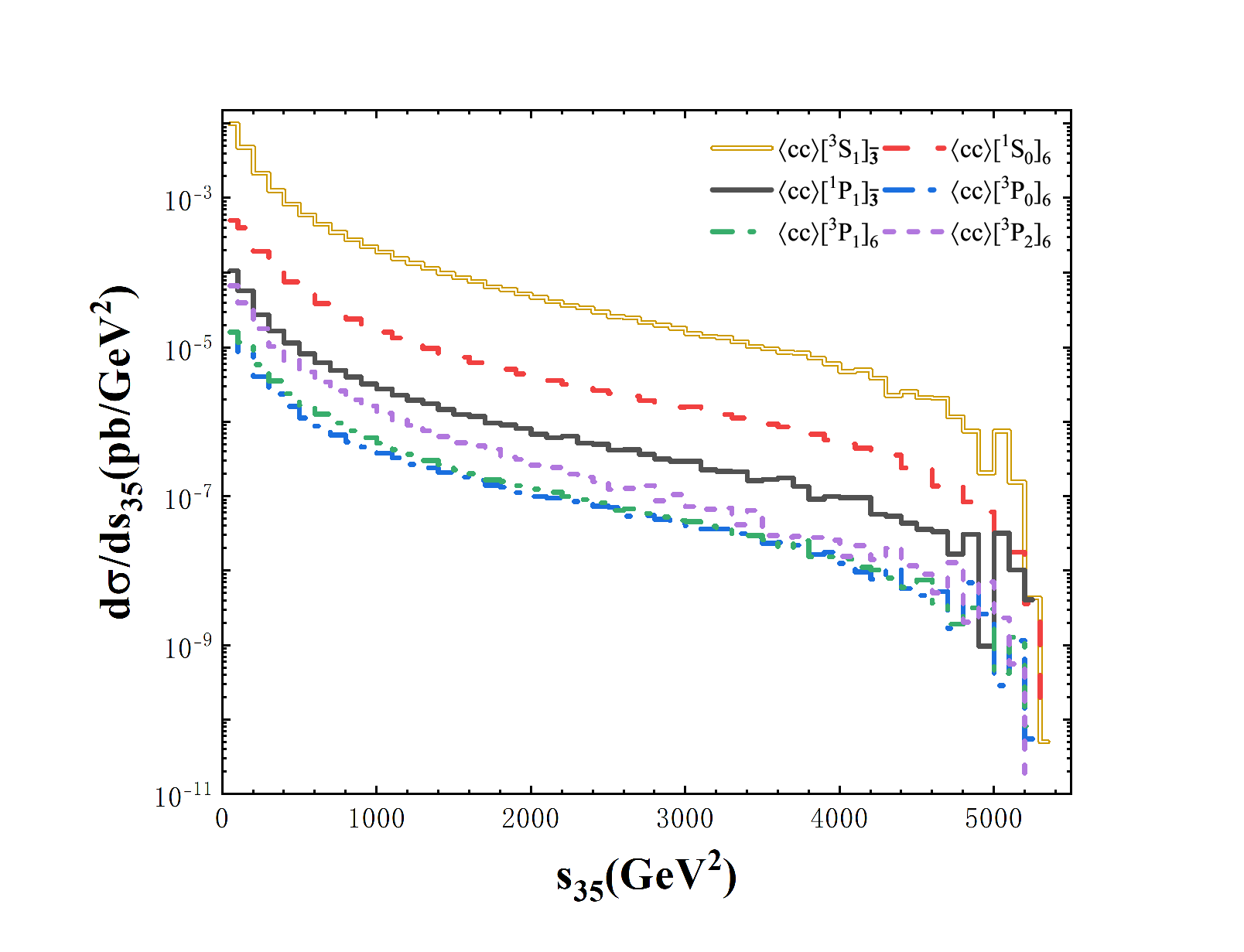}
\hspace{-0.50in}
    \includegraphics[scale=0.22]{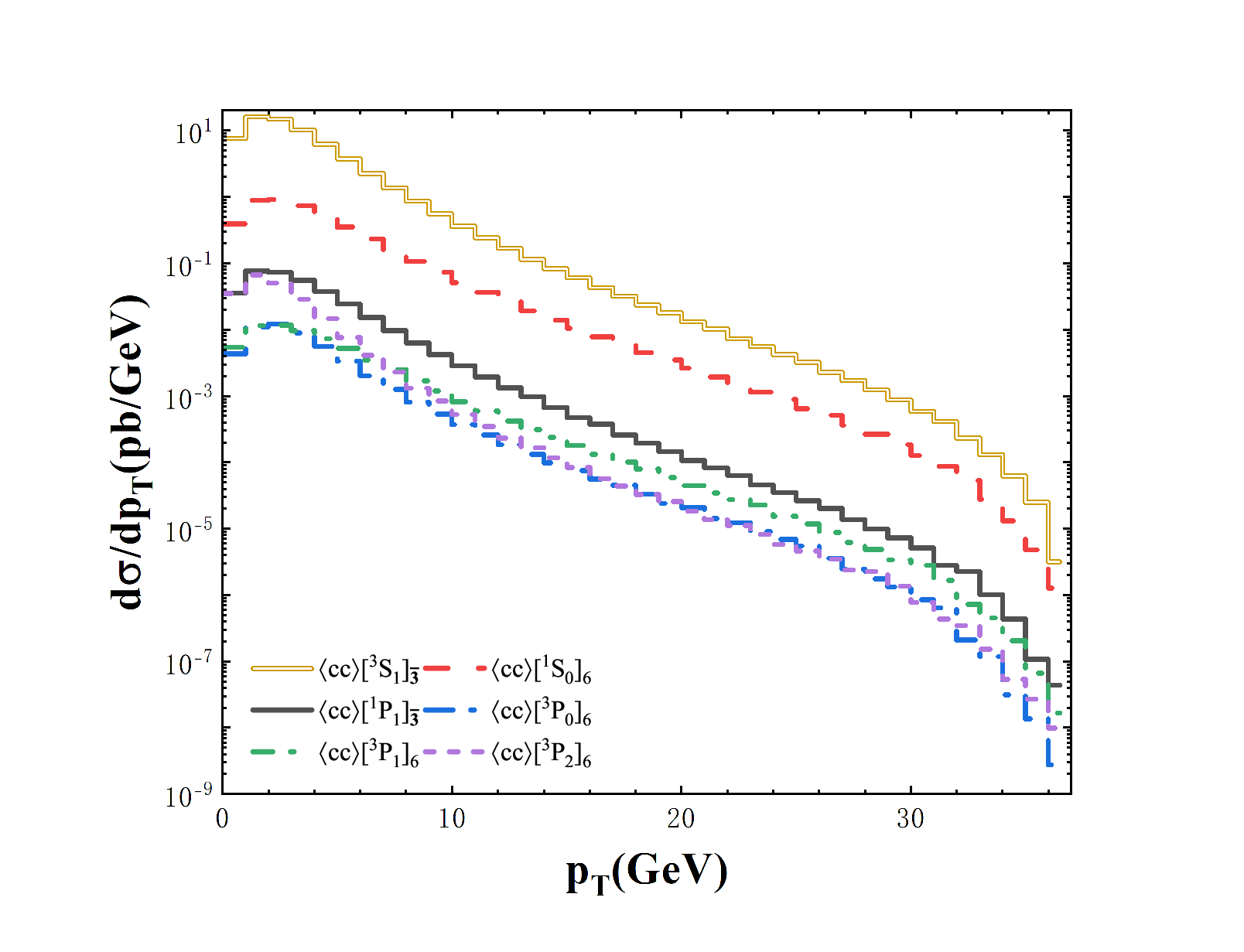}
\caption{Differential distributions for the photoproduction of $\Xi_{cc}$ at $\sqrt{s}=91$ GeV.} \label{ccdist}
\end{figure}

\begin{figure}
\centering
\hspace{-0.50in}
    \includegraphics[scale=0.22]{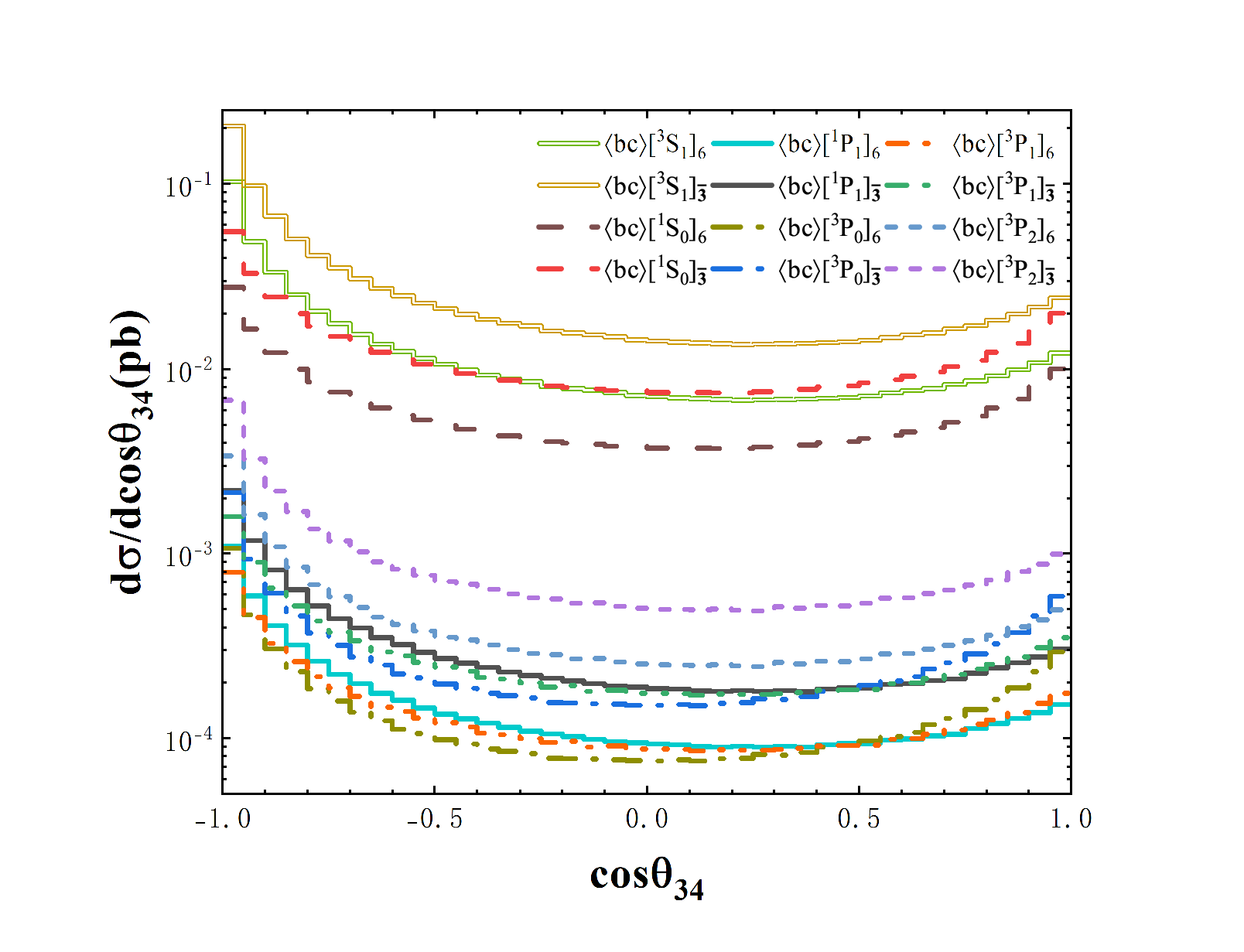}
  \hspace{-0.50in}
    \includegraphics[scale=0.22]{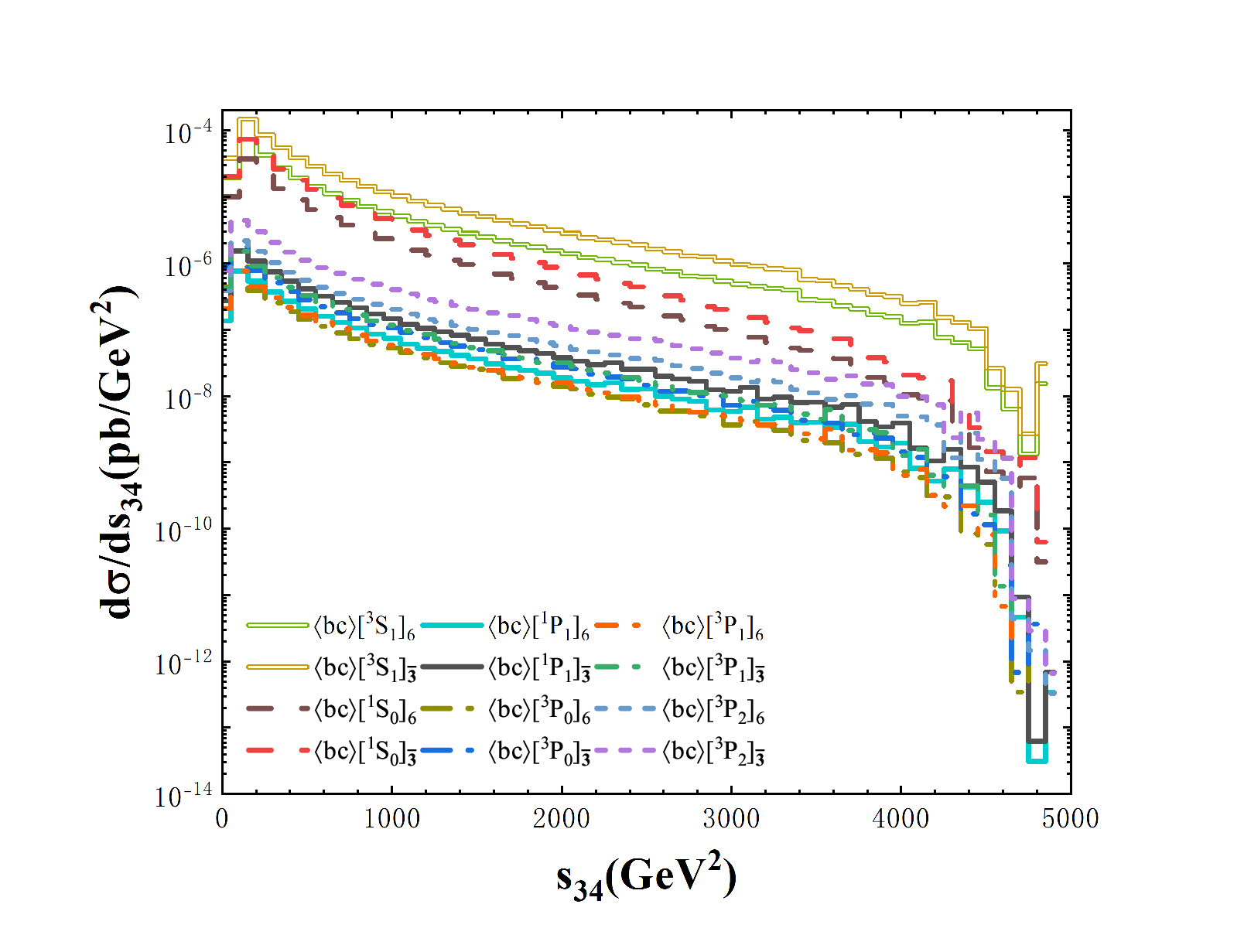}
  \hspace{-0.50in}
    \includegraphics[scale=0.22]{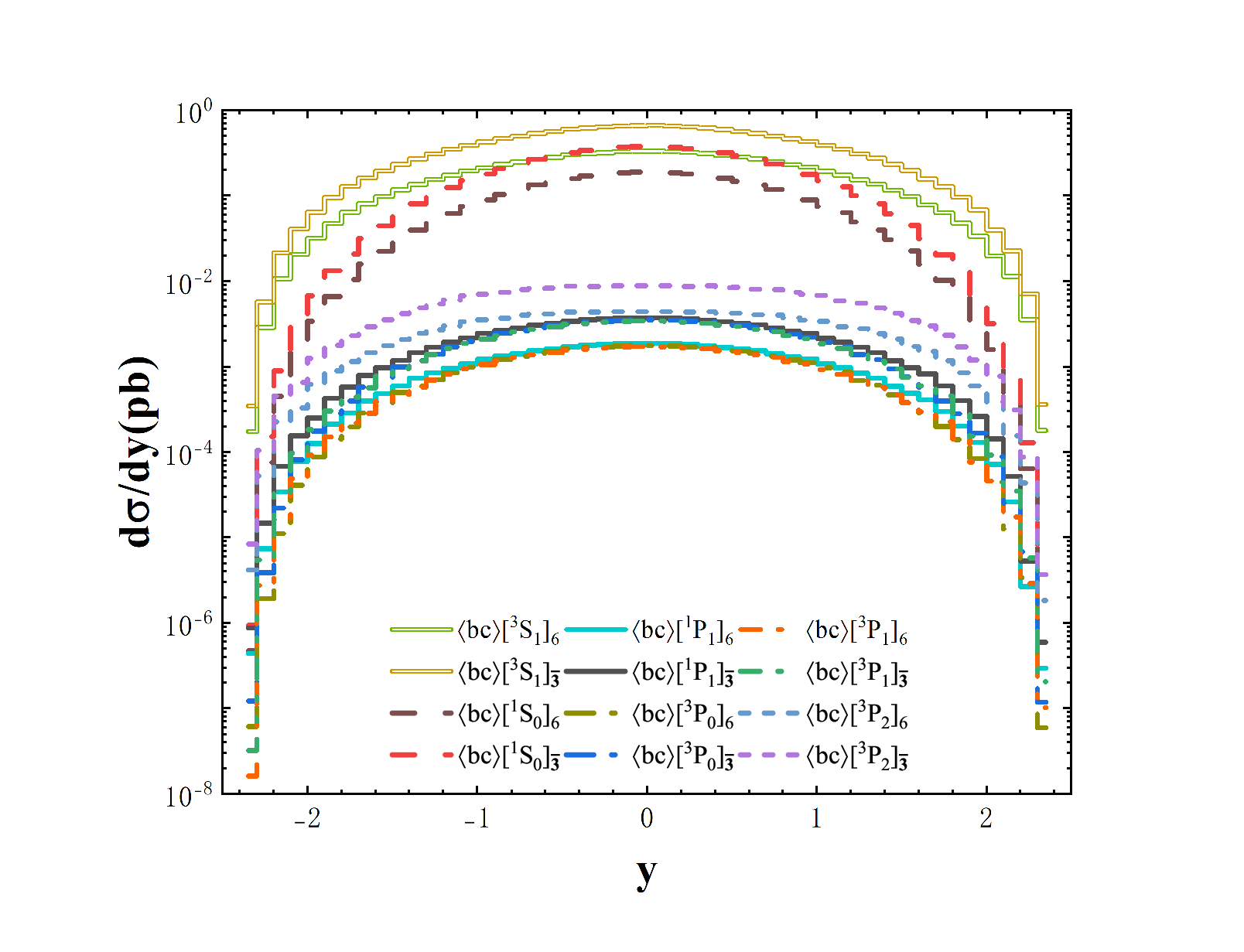}\\
\hspace{-0.50in}
    \includegraphics[scale=0.22]{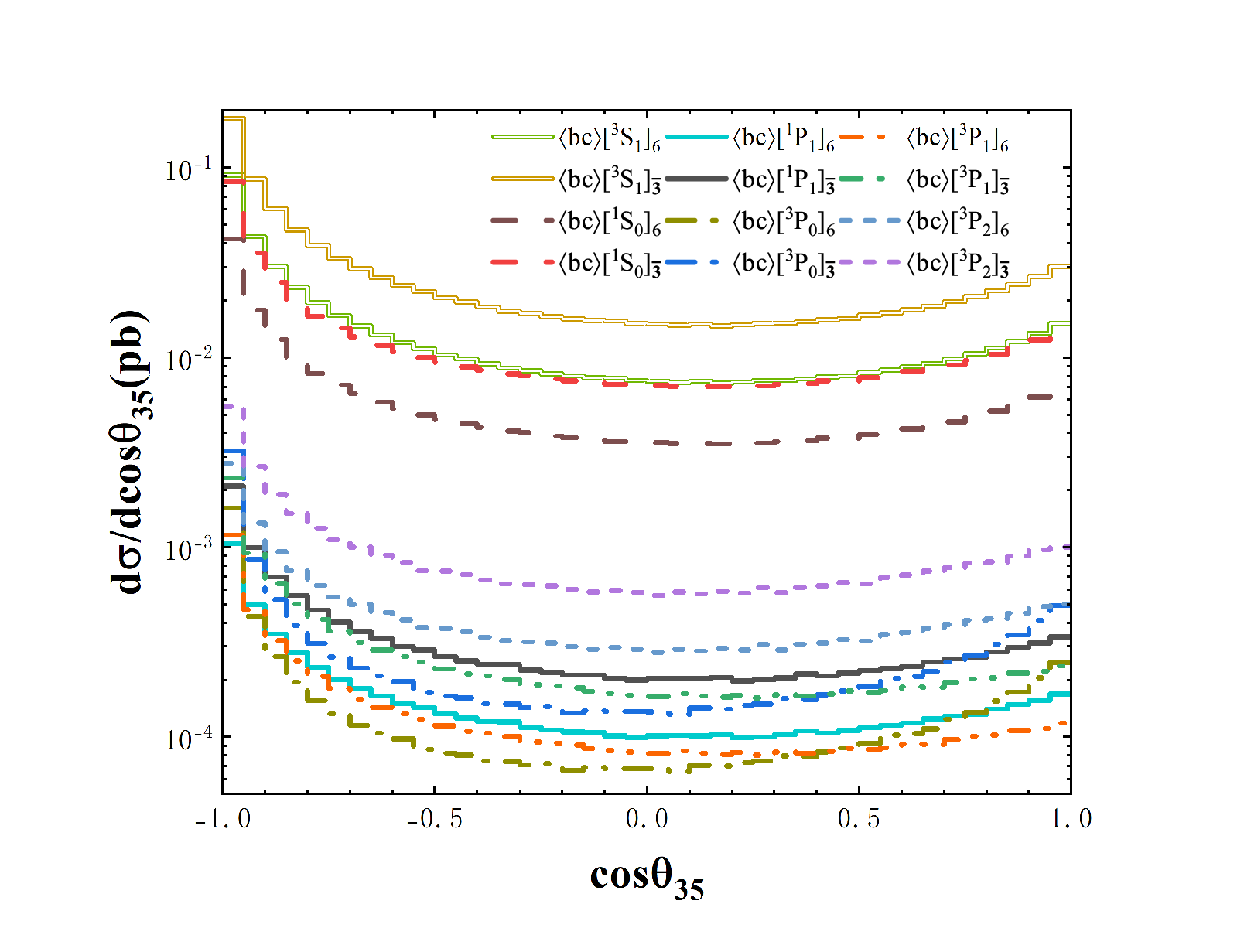}
\hspace{-0.50in}
    \includegraphics[scale=0.22]{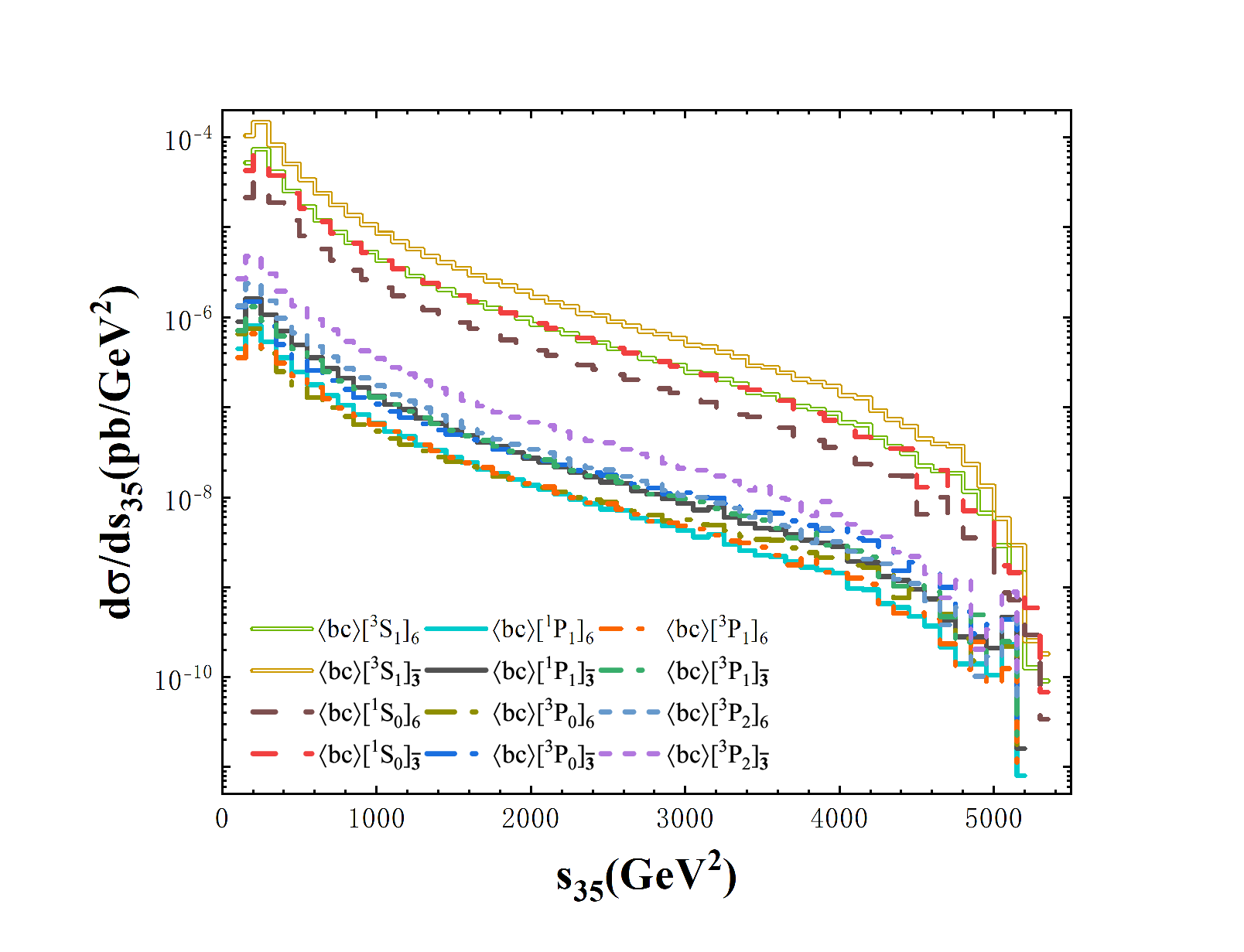}
\hspace{-0.50in}
    \includegraphics[scale=0.22]{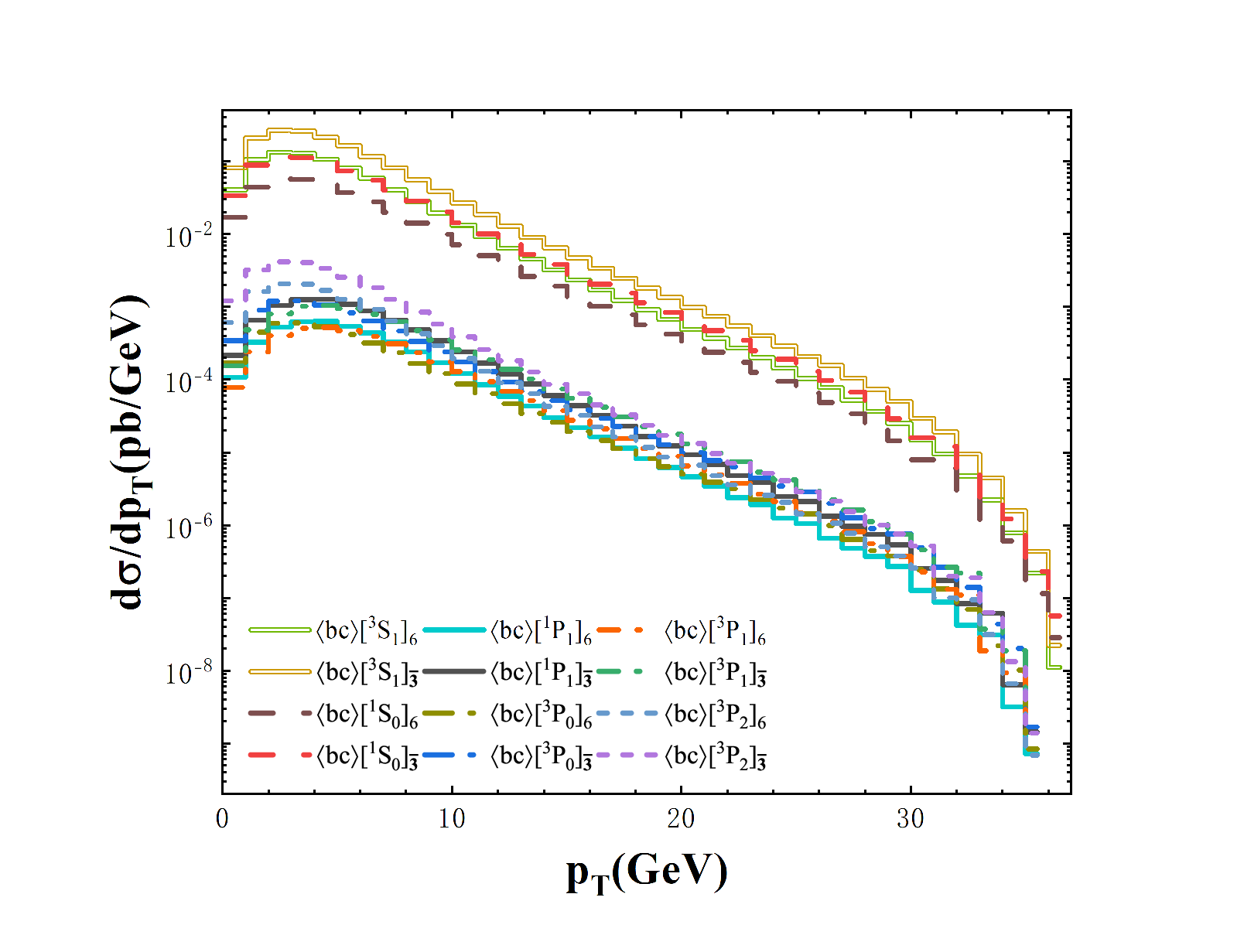}
\caption{Differential distributions for the photoproduction of $\Xi_{bc}$ at $\sqrt{s}=91$ GeV.} \label{bcdist}
\end{figure}

\begin{figure}
\centering
\hspace{-0.50in}
    \includegraphics[scale=0.22]{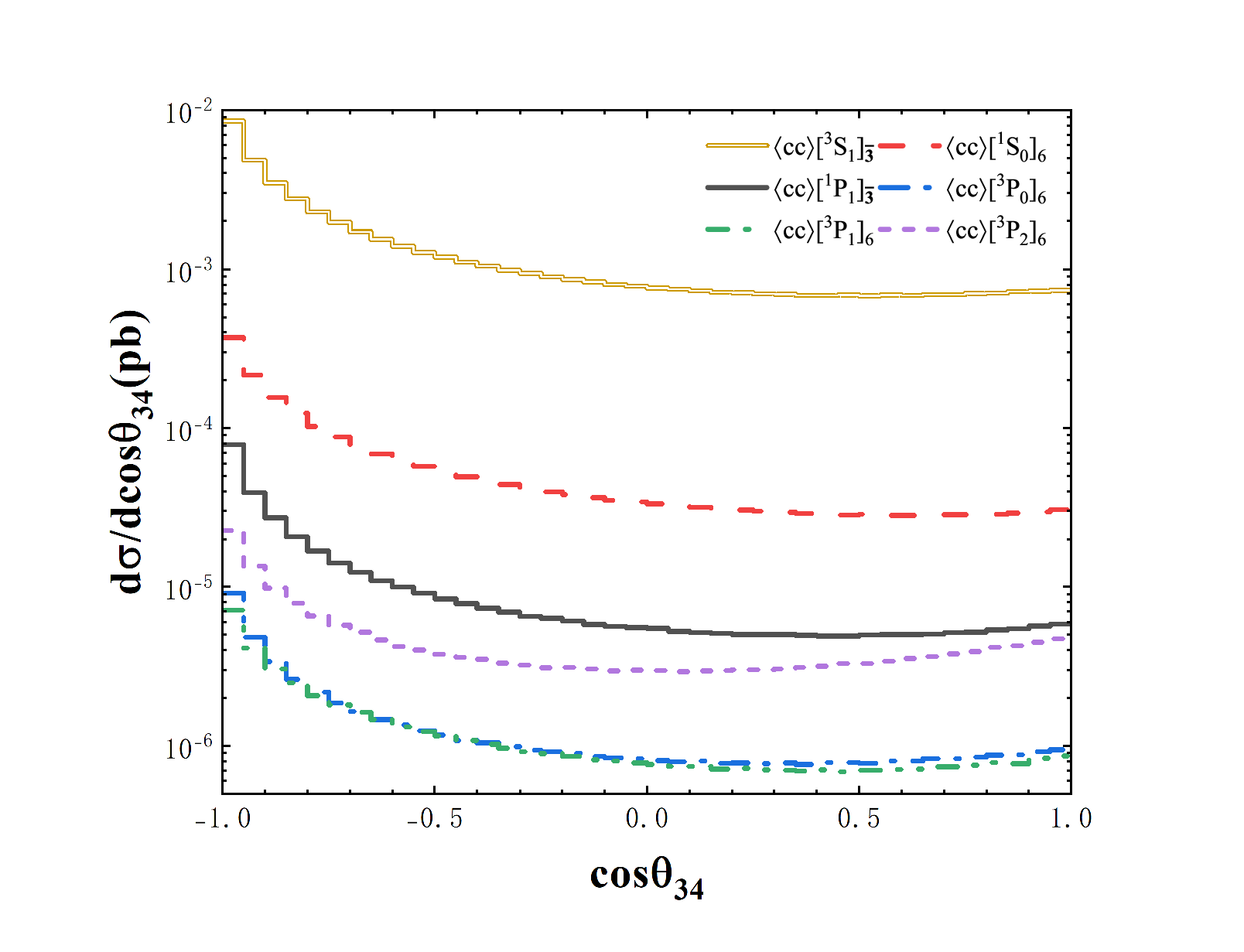}
  \hspace{-0.50in}
    \includegraphics[scale=0.22]{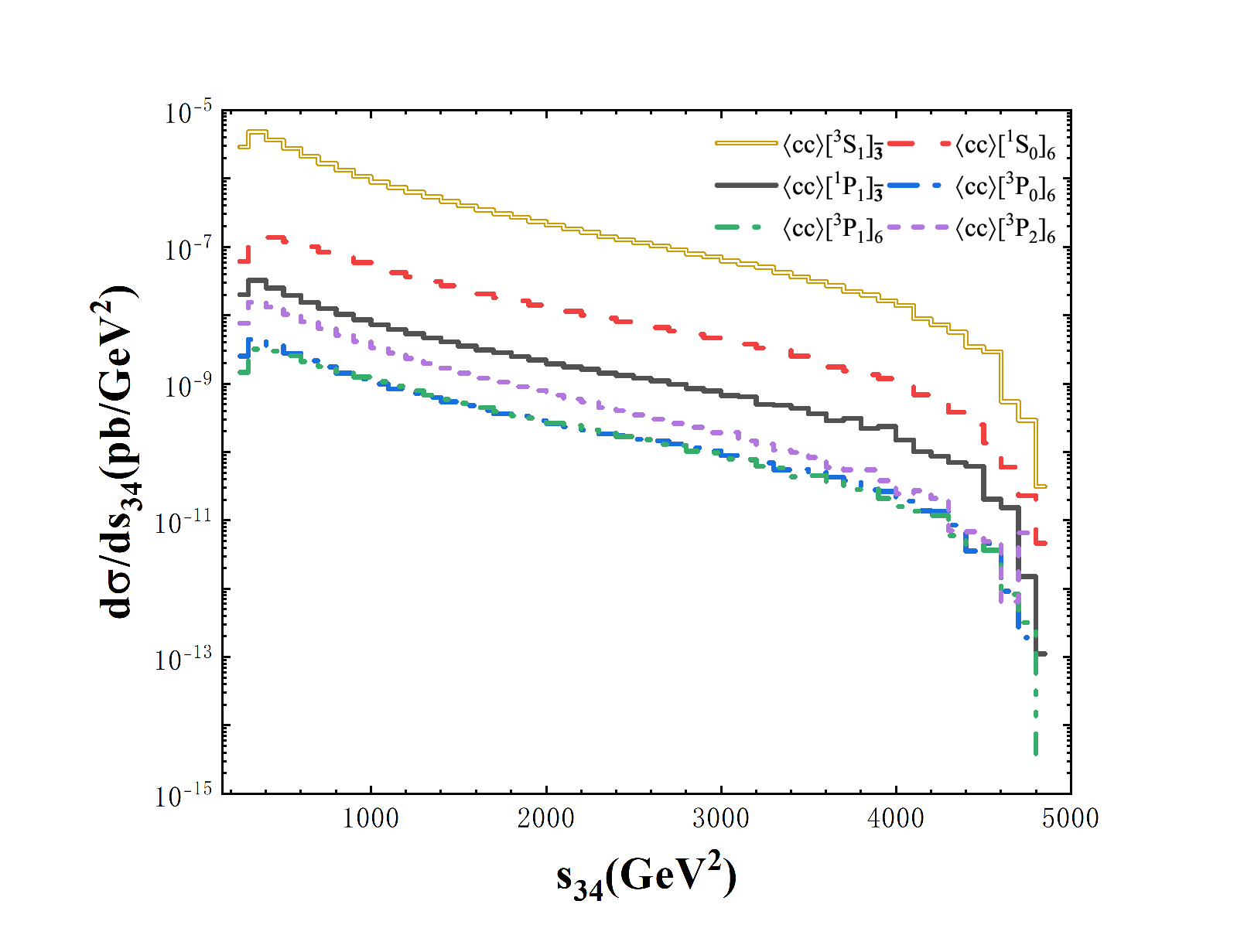}
  \hspace{-0.50in}
    \includegraphics[scale=0.22]{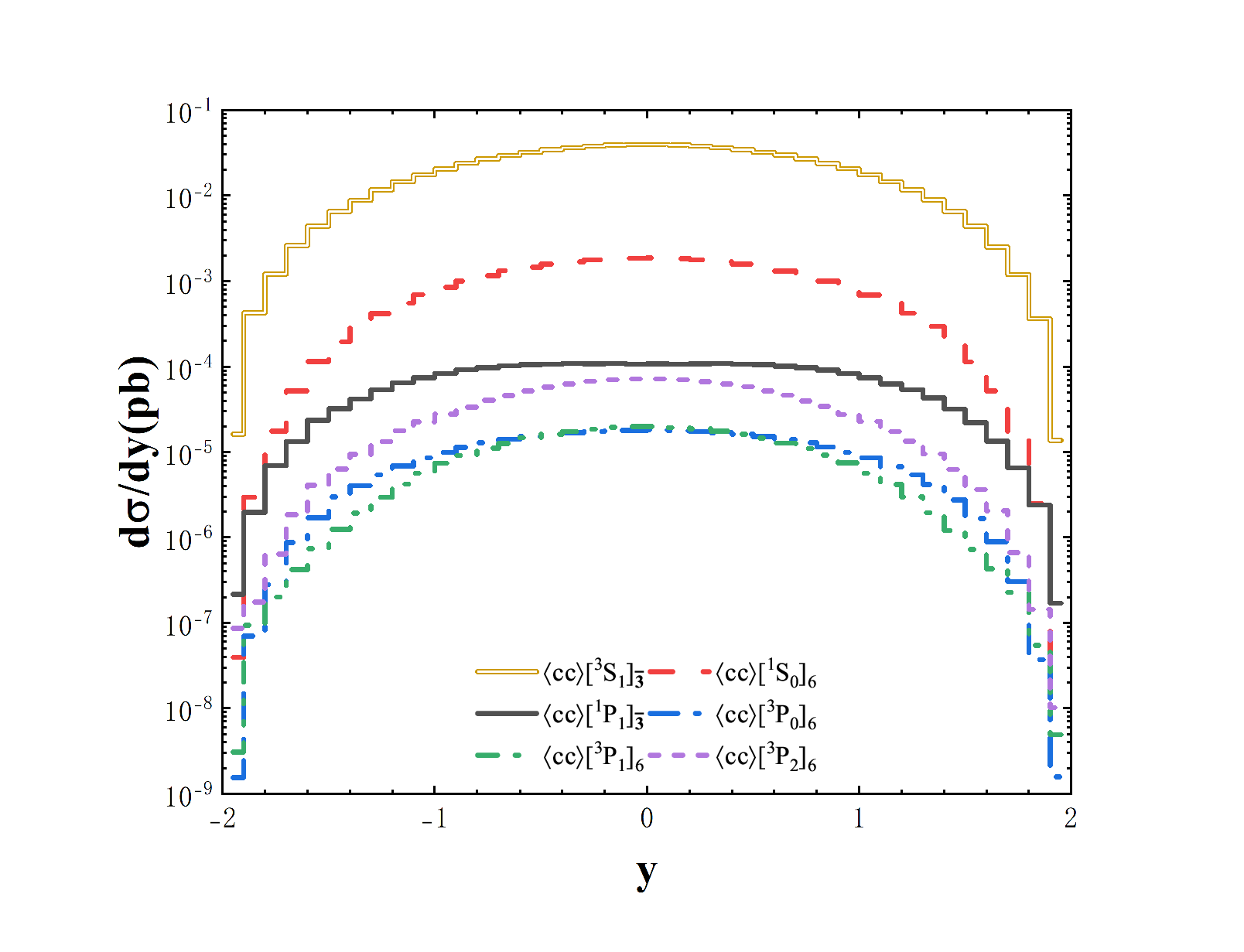}\\
\hspace{-0.50in}
    \includegraphics[scale=0.22]{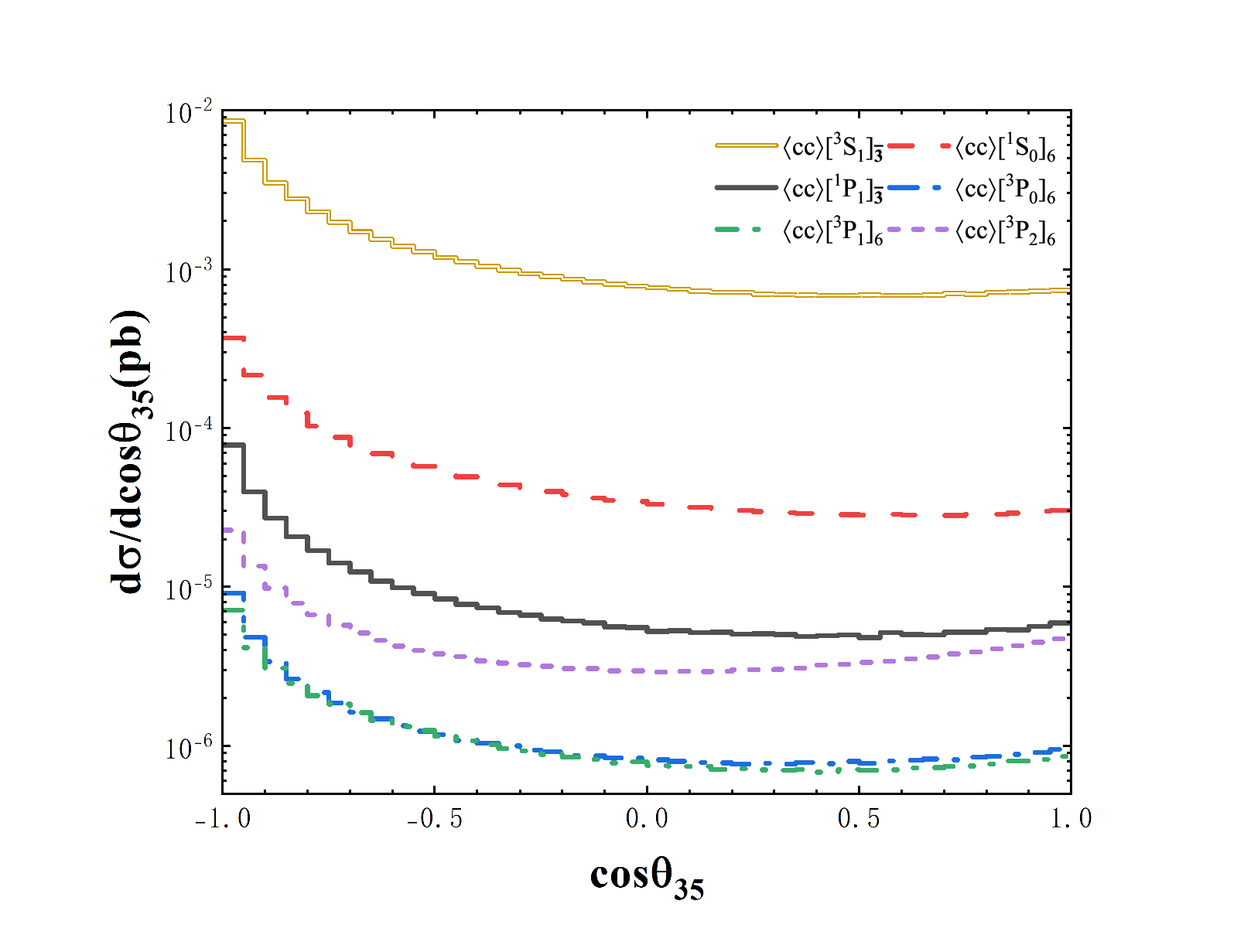}
\hspace{-0.50in}
    \includegraphics[scale=0.22]{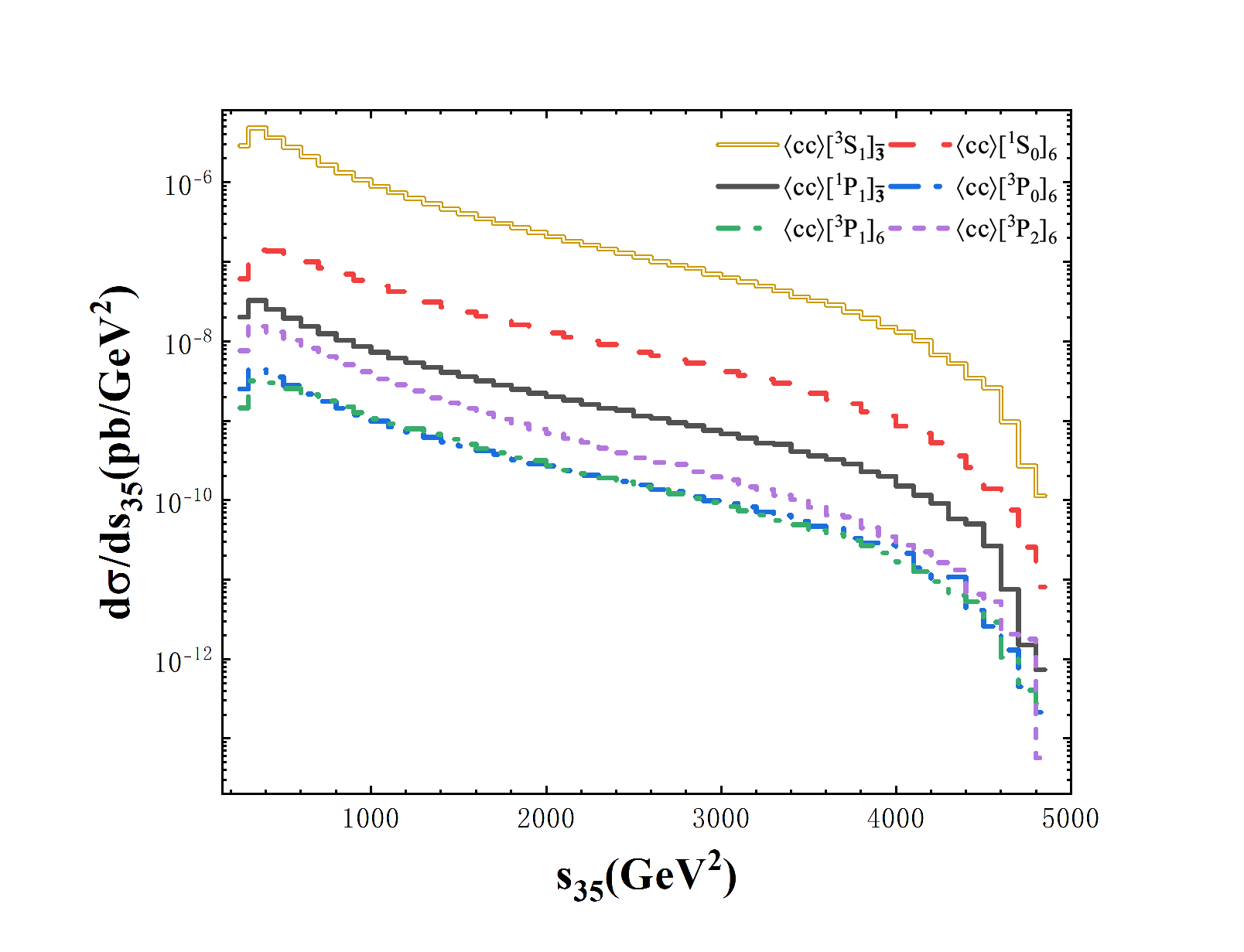}
\hspace{-0.50in}
    \includegraphics[scale=0.22]{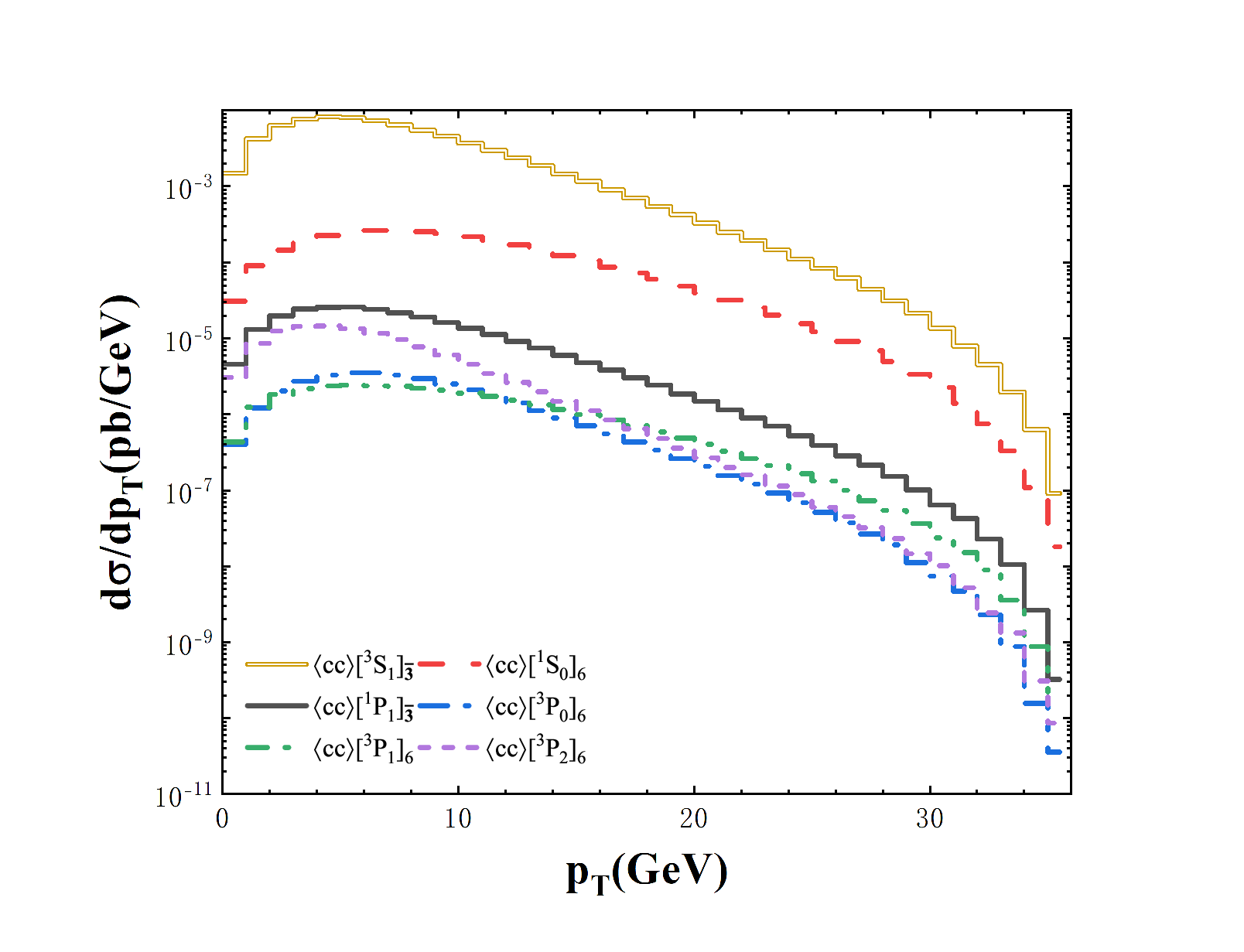}
\caption{Differential distributions for the photoproduction of $\Xi_{bb}$ at $\sqrt{s}=91$ GeV.} \label{bbdist}
\end{figure}

To show the dynamical behavior of the produced $\Xi_{QQ^{\prime}}$ baryon from different intermediate diquark state $\langle QQ^{\prime} \rangle[n]$ and be helpful for the experiments measurements, the differential cross sections, such as the angular ($\cos\theta_{34}$ and $\cos\theta_{35}$), invariant mass ($s_{34}$ and $s_{35}$), transverse momentum ($p_T$), and rapidity ($y$) distributions,
of $\Xi_{cc}$, $\Xi_{bc}$ and $\Xi_{bb}$ are presented in Figs.~\ref{ccdist}-\ref{bbdist}, respectively.
Here $\theta_{ij}$ is the angle between the outgoing three momenta $\vec{p}_i$ and $\vec{p}_j$ in the center-of-mass system. The definition of invariant mass $s_{ij}= (p_i+p_j)^2$. 
The different colors and linearity in the Figs.~\ref{ccdist}-\ref{bbdist} represent the differential distributions of intermediate diquark configures.

It can be seen from the transverse momentum distribution in Figs.~\ref{ccdist}-\ref{bbdist} that there is a sharply peak in small $p_T$ region, around several GeV, and then a logarithmic decline for the photoproduction of $\Xi_{cc}$, $\Xi_{bc}$. For $\Xi_{bb}$ baryon, however, the peak is relatively flat but the general trend is consistent with $\Xi_{cc}$, $\Xi_{bc}$.
Throughout the whole $p_T$ region, the trend of different intermediate diquark states is generally similar, and the contribution of $[^3S_1]_{\mathbf{\bar 3}}$ configuration is consistently dominant. The rapidity $y$ distributions of $\Xi_{QQ^{\prime}}$ baryons are mainly in the range of -2.5 to 2.5, and is reduced to the range of -2.3 to 2.3 (-2.0 to 2.0) for the photoproduction of $\Xi_{bc}$ ($\Xi_{bb}$). The invariant mass distributions $s_{34}$ and $s_{35}$ are monotonically decreasing. This means that the contribution is greatest when the baryon is moving back to back with two anti-charm quarks. The same conclusion can also be confirmed by the angular $\cos\theta_{34}$ and $\cos\theta_{35}$ distributions, when $\cos\theta_{34}$ and $\cos\theta_{35}$ are equal to -1, the maximum cross section can be achieved.

\subsection{Theoretical uncertainty}

In this subsection, we focus on the analysis of two main sources of theoretical uncertainty in the photoproduction of $\Xi_{cc}$, $\Xi_{bc}$, and $\Xi_{bb}$ baryons, involoving heavy quark masses $m_c$ and $m_b$ and the renormalization energy scale $\mu_0$. Note that when calculating the theoretical uncertainty arising from a variable, the other parameters remain at their central values.

Firstly, we analyze the theoretical uncertainty of the cross sections induced by varying $m_c=1.8\pm 0.3$ GeV, and the results are listed in Tables~\ref{mc_unser}-\ref{Xibc_muncer} corresponding to the photoproduction of $\Xi_{cc}$ and $\Xi_{bc}$ baryons at $\sqrt{s}=91$ GeV. The uncertainty caused by changing $m_b=5.1\pm 0.3$ GeV are shown in Tables \ref{Xibc_muncer}-\ref{mb_uncer}, corresponding to the photoproduction of $\Xi_{bc}$ and $\Xi_{bb}$ baryons, respectively.

From three Tables~\ref{mc_unser}-\ref{mb_uncer}, the dependence of the cross section from each intermediate diquark state on heavy quark mass is significant. The cross section tends to decrease with the increase of $m_c$ and $m_b$ mainly due to the inhibition of phase space. The dependence of $\Xi_{cc}$ baryon on $m_c$ is obviously greater than that of $\Xi_{bc}$ baryon. Similarly, the $\Xi_{bb}$ baryon is significantly more dependent on $m_b$ than $\Xi_{bc}$ baryon. We numerically analyze the theoretical uncertainty arising from heavy quark mass by summing the contributions of all intermediate diquark states (total). Specifically, the total cross section of $\Xi_{cc}$ baryon can vary by $-53.46\%\sim +145.31\%$ for $m_c=1.8\pm 0.3$ GeV in Table~\ref{mc_unser}; the total cross section of $\Xi_{bc}$ baryons can vary by $-31.53\%\sim +57.26\%$ for $m_c=1.8\pm 0.3$ and $-16.13\%\sim+18.55\%$ for $m_b=5.1\pm 0.3$ GeV in Tables~\ref{Xibc_muncer}; and the total cross section of $\Xi_{bb}$ baryons can vary by
$-27.82\%\sim+39.79\%$ for $m_b=5.1\pm 0.3$ GeV in Tables~\ref{mb_uncer}.

\begin{table}[htb]
\begin{center}
\caption{Theoretical uncertainty of the cross sections (in unit: pb) by varying $m_c = 1.8 \pm 0.3$ GeV for the photoproduction of $\Xi_{cc}$ at $\sqrt{s}=91$ GeV.} \vspace{0.5cm}
\begin{tabular}{|c|c|c|c|}
\hline

 State & $m_c = 1.5$ & $m_c = 1.8$ & $m_c = 2.1$\\
\hline
$[^1S_0]_{\mathbf 6}$               & $4.03 ~\times ~10^{-1}$ & $1.58 ~\times ~10^{-1}$ & $7.07 ~\times ~10^{-2}$\\
\hline
$[^3S_1]_{\mathbf{\bar 3}}$ &         $5.42 $         &         $2.24 $         &        $1.05 $        \\
\hline
$[^1P_1]_{\mathbf{\bar 3}}$ & $9.53 ~\times ~10^{-2}$ & $2.70 ~\times ~10^{-2}$ & $9.27 ~\times ~10^{-3}$\\
\hline
$[^3P_0]_{\mathbf 6}$               & $1.42 ~\times ~10^{-2}$ & $3.99 ~\times ~10^{-3}$ & $1.36 ~\times ~10^{-3}$\\
\hline
$[^3P_1]_{\mathbf 6}$               & $1.79 ~\times ~10^{-2}$ & $4.89 ~\times ~10^{-3}$ & $1.62 ~\times ~10^{-3}$\\
\hline
$[^3P_2]_{\mathbf 6}$               & $6.04 ~\times ~10^{-2}$ & $1.66 ~\times ~10^{-2}$ & $5.50 ~\times ~10^{-3}$\\
\hline
total               &  6.01 & 2.45 & 1.14 \\
\hline
\end{tabular}
\label{mc_unser}
\end{center}
\end{table}

\begin{table}[htb]
\begin{center}
\caption{Theoretical uncertainty of the cross sections (in unit: pb) by varying $m_c = 1.8 \pm 0.3$ GeV or $m_b = 5.1 \pm 0.3$ GeV respectively for the photoproduction of $\Xi_{bc}$ at $\sqrt{s}=91$ GeV.} \vspace{0.5cm}
\begin{tabular}{|c|c|c|c|c|c|}
\hline

 State & $m_c = 1.5$ & $m_{c,b} = 1.8, 5.1$  & $m_c = 2.1$ & $m_b = 4.8$ & $m_b = 5.4$ \\
  \hline
$[^1S_0]_{\mathbf 6}$             & $1.87 \times 10^{-2}$ & $1.23 \times 10^{-2}$  &  $8.68 \times 10^{-3}$ & $1.47 \times 10^{-2}$ & $1.05 \times 10^{-2}$\\
\hline
$[^1S_0]_{\mathbf{\bar 3}}$ & $3.75 \times 10^{-2}$ & $2.47 \times 10^{-2}$  &  $1.74 \times 10^{-2}$ &$2.94 \times 10^{-2}$ & $2.09 \times 10^{-2}$\\
\hline
$[^3S_1]_{\mathbf 6}$               & $4.26 \times 10^{-2}$ & $2.73 \times 10^{-2}$  &  $1.87 \times 10^{-2}$ & $3.21 \times 10^{-2}$ & $2.34 \times 10^{-2}$\\
\hline
$[^3S_1]_{\mathbf{\bar 3}}$ & $8.52 \times 10^{-2}$ & $5.47 \times 10^{-2}$  &  $3.75 \times 10^{-2}$ & $6.42 \times 10^{-2}$ & $4.68 \times 10^{-2}$\\
\hline
$[^1P_1]_{\mathbf 6}$               & $7.66 \times 10^{-4}$ & $3.39 \times 10^{-4}$  &  $1.71 \times 10^{-4}$ & $4.04 \times 10^{-4}$ & $2.87 \times 10^{-4}$\\
\hline
$[^1P_1]_{\mathbf{\bar 3}}$ & $1.53 \times 10^{-3}$ & $6.78 \times 10^{-4}$  &  $3.43 \times 10^{-4}$ & $8.09 \times 10^{-4}$ &  $5.73 \times 10^{-4}$\\
\hline
$[^3P_0]_{\mathbf 6}$               & $1.96 \times 10^{-4}$  & $1.02 \times 10^{-4}$ & $5.96 \times 10^{-5}$  & $1.28 \times 10^{-4}$ & $8.27 \times 10^{-5}$ \\
\hline
$[^3P_0]_{\mathbf{\bar 3}}$ & $3.92 \times 10^{-4}$  & $2.04 \times 10^{-4}$ & $1.19 \times 10^{-4}$  & $2.56 \times 10^{-4}$  &  $1.65 \times 10^{-4}$\\
\hline
$[^3P_1]_{\mathbf 6}$               & $6.24 \times 10^{-4}$  & $2.97 \times 10^{-4}$ & $1.61 \times 10^{-4}$ & $3.62 \times 10^{-4}$  & $2.46 \times 10^{-4}$ \\
\hline
$[^3P_1]_{\mathbf{\bar 3}}$ & $1.25 \times 10^{-3}$  & $5.95 \times 10^{-4}$ & $3.22 \times 10^{-4}$ & $7.25 \times 10^{-4}$  & $4.93 \times 10^{-4}$ \\
\hline
$[^3P_2]_{\mathbf 6}$       & $2.09 \times 10^{-3}$  & $9.49 \times 10^{-4}$ & $4.89 \times 10^{-4}$  & $1.14 \times 10^{-3}$  &  $7.97 \times 10^{-4}$\\
\hline
$[^3P_2]_{\mathbf{\bar 3}}$ & $4.18 \times 10^{-3}$  & $1.90 \times 10^{-3}$ & $9.78 \times 10^{-4}$ & $2.28 \times 10^{-3}$  &  $1.59 \times 10^{-3}$ \\
\hline
total &  $1.95\times 10^{-1}$ & $1.24\times 10^{-1}$ & $8.49\times10^{-2}$ & $1.47 \times 10^{-1}$ & $1.04 \times 10^{-1}$\\
\hline
\end{tabular}
\label{Xibc_muncer}
\end{center}
\end{table}

\begin{table}[htb]
\begin{center}
\caption{Theoretical uncertainty of the cross sections (in unit: pb) by varying $m_b = 5.1 \pm 0.3$ GeV for the photoproduction of $\Xi_{bb}$ at $\sqrt{s}=91$ GeV.} \vspace{0.5cm}
\begin{tabular}{|c|c|c|c|}
\hline

 State & $m_b = 4.8$ & $m_b = 5.1$ & $m_b = 5.4$\\
\hline
$[^1S_0]_{\mathbf6}$               & $1.70 ~\times ~10^{-4}$ & $1.17 ~\times ~10^{-4}$ & $8.18 ~\times ~10^{-5}$\\
\hline
$[^3S_1]_{\mathbf{\bar 3}}$ & $3.74 ~\times ~10^{-3}$ & $2.68 ~\times ~10^{-3}$ & $1.95 ~\times ~10^{-3}$\\
\hline
$[^1P_1]_{\mathbf{\bar 3}}$ & $3.27 ~\times ~10^{-5}$ & $2.05 ~\times ~10^{-5}$ & $1.32 ~\times ~10^{-5}$\\
\hline
$[^3P_0]_{\mathbf6}$               & $4.47 ~\times ~10^{-6}$ & $2.79 ~\times ~10^{-6}$ & $1.77 ~\times ~10^{-6}$\\
\hline
$[^3P_1]_{\mathbf6}$               & $4.17 ~\times ~10^{-6}$ & $2.53 ~\times ~10^{-6}$ & $1.57 ~\times ~10^{-6}$\\
\hline
$[^3P_2]_{\mathbf6}$               & $1.53 ~\times ~10^{-5}$ & $9.36 ~\times ~10^{-6}$ & $5.86 ~\times ~10^{-6}$\\
\hline
total    & $3.97 ~\times ~10^{-3}$  & $2.84 ~\times ~10^{-3}$ & $2.05 ~\times ~10^{-3}$   \\
\hline
\end{tabular}
\label{mb_uncer}
\end{center}
\end{table}

Then the theoretical uncertainty of the cross sections is analyzed by varying the renormalization energy scale $\mu$ from $\frac{1}{2}\mu_0$ to $2\mu_0$ with $\mu_0=\sqrt{M_{\Xi_{QQ^{\prime}}}^2+p_{T}^2}$ for the photoproduction of $\Xi_{QQ^{\prime}}$ at $\sqrt{s}=91$ GeV, and the corresponding results are presented in Tables~\ref{mc_as_uncer}-\ref{mb_asuncer}. At leading-order, the square of the amplitude for the photoproduction of $\Xi_{QQ^{\prime}}$ baryon is proportional to the strong coupling $\alpha_s^2$, which is running with the energy scale $\mu$ and shows a strong dependence on $\mu$.

Tables~\ref{mc_as_uncer}-\ref{mb_asuncer} also reflect this theoretical uncertainty of the energy scale $\mu$ numerically. 
By changing the energy scale $\mu=\frac{1}{2}\mu_0$ and 2$\mu_0$, the total cross section will change by $-30.17\%\sim+67.36\%$, $-24.03\%\sim+54.03\%$, and $-25.09\%\sim+40.28\%$
for the photoproduction of $\Xi_{cc}$, $\Xi_{bc}$, and $\Xi_{bb}$, respectively. With the help of subsequent higher-order calculations, the energy scale uncertainty will be greatly reduced until more accurate numerical results are obtained. Nevertheless, the photoproduced events of doubly heavy baryons $\Xi_{QQ^{\prime}}$ remain appreciable at CEPC and FCC-ee, and further experimental testing of the doubly heavy baryons in future will be expected.

\begin{table}[htb]
\begin{center}
\caption{Theoretical uncertainty of the cross sections (in unit: pb) by varying $\mu$ = $\frac{1}{2}\mu_0$, $\mu_0$ and 2$\mu_0$ for the photoproduction of $\Xi_{cc}$ at $\sqrt{s}=91$ GeV, where $\mu_0=\sqrt{M_{\Xi_{cc}}^2+p_{T}^2}$.} \vspace{0.5cm}
\begin{tabular}{|c|c|c|c|}
\hline

 State & 0.5$\mu_0$ & $\mu_0$ & 2$\mu_0$\\
\hline
$[^1S_0]_{\mathbf 6}$               & $2.60 ~\times ~10^{-1}$ & $1.58 ~\times ~10^{-1}$ & $1.09 ~\times ~10^{-1}$\\
\hline
$[^3S_1]_{\mathbf{\bar 3}}$ &         $3.75 $         &         $2.24 $         &        $1.54 $        \\
\hline
$[^1P_1]_{\mathbf{\bar 3}}$ & $4.50 ~\times ~10^{-2}$ & $2.70 ~\times ~10^{-2}$ & $1.87 ~\times ~10^{-2}$\\
\hline
$[^3P_0]_{\mathbf 6}$               & $6.64 ~\times ~10^{-3}$ & $3.99 ~\times ~10^{-3}$ & $2.75 ~\times ~10^{-3}$\\
\hline
$[^3P_1]_{\mathbf 6}$               & $8.03 ~\times ~10^{-3}$ & $4.89 ~\times ~10^{-3}$ & $3.41 ~\times ~10^{-3}$\\
\hline
$[^3P_2]_{\mathbf 6}$               & $2.82 ~\times ~10^{-2}$ & $1.66 ~\times ~10^{-2}$ & $1.13 ~\times ~10^{-2}$\\
\hline
total      & 4.05 & 2.42  & 1.69 \\
\hline
\end{tabular}
\label{mc_as_uncer}
\end{center}
\end{table}

\begin{table}[htb]
\begin{center}
\caption{Theoretical uncertainty of the cross sections (in unit: pb) by varying $\mu$ = $\frac{1}{2}\mu_0$, $\mu_0$ and 2$\mu_0$ for the photoproduction of $\Xi_{bc}$ at $\sqrt{s}=91$ GeV, where $\mu_0=\sqrt{M_{\Xi_{bc}}^2+p_{T}^2}$.} \vspace{0.5cm}
\begin{tabular}{|c|c|c|c|}
\hline

 State & 0.5$\mu_0$ & $\mu_0$ & 2$\mu_0$\\
 \hline
$[^1S_0]_{6}$               & $1.83 ~\times ~10^{-2}$ & $1.23 ~\times ~10^{-2}$ & $9.02 ~\times ~10^{-3}$ \\
\hline
$[^1S_0]_{\mathbf{\bar 3}}$ & $3.65 ~\times ~10^{-2}$ & $2.47 ~\times ~10^{-2}$ & $1.80 ~\times ~10^{-2}$ \\
\hline
$[^3S_1]_{6}$               & $4.05 ~\times ~10^{-2}$ & $2.73 ~\times ~10^{-2}$ & $1.99 ~\times ~10^{-2}$\\
\hline
$[^3S_1]_{\mathbf{\bar 3}}$ & $8.10 ~\times ~10^{-2}$ & $5.47 ~\times ~10^{-2}$ & $3.99 ~\times ~10^{-2}$\\
\hline
$[^1P_1]_{6}$               & $9.96 ~\times ~10^{-4}$ & $3.39 ~\times ~10^{-4}$ & $4.97 ~\times ~10^{-4}$\\
\hline
$[^1P_1]_{\mathbf{\bar 3}}$ & $1.99 ~\times ~10^{-3}$ & $6.78 ~\times ~10^{-4}$ & $9.94 ~\times ~10^{-4}$\\
\hline
$[^3P_0]_{6}$               & $3.02 ~\times ~10^{-4}$ & $1.02 ~\times ~10^{-4}$ & $1.50 ~\times ~10^{-4}$\\
\hline
$[^3P_0]_{\mathbf{\bar 3}}$ & $6.04 ~\times ~10^{-4}$ & $2.04 ~\times ~10^{-4}$ & $2.99 ~\times ~10^{-4}$\\
\hline
$[^3P_1]_{6}$               & $8.70 ~\times ~10^{-4}$ & $2.97 ~\times ~10^{-4}$ & $4.37 ~\times ~10^{-4}$\\
\hline
$[^3P_1]_{\mathbf{\bar 3}}$ & $1.74 ~\times ~10^{-3}$ & $5.95 ~\times ~10^{-4}$ & $8.73 ~\times ~10^{-4}$\\
\hline
$[^3P_2]_{6}$               & $2.82 ~\times ~10^{-3}$ & $9.49 ~\times ~10^{-4}$ & $1.39 ~\times ~10^{-3}$\\
\hline
$[^3P_2]_{\mathbf{\bar 3}}$ & $5.64 ~\times ~10^{-3}$ & $1.90 ~\times ~10^{-3}$ & $2.77 ~\times ~10^{-3}$\\
\hline
total  & $1.91 ~\times ~10^{-1}$ & $1.24 ~\times ~10^{-1}$ & $9.42 ~\times ~10^{-2}$  \\
\hline
\end{tabular}
\label{Xibc_asuncer}
\end{center}
\end{table}

\begin{table}[htb]
\begin{center}
\caption{Theoretical uncertainty of the cross sections (in unit: pb) by varying $\mu$ = $\frac{1}{2}\mu_0$, $\mu_0$ and 2$\mu_0$ for the photoproduction of $\Xi_{bb}$ at $\sqrt{s}=91$ GeV, where $\mu_0=\sqrt{M_{\Xi_{bb}}^2+p_{T}^2}$.} \vspace{0.5cm}
\begin{tabular}{|c|c|c|c|}
\hline

 State & 0.5$\mu_0$ & $\mu_0$ & 2$\mu_0$\\
\hline
$[^1S_0]_{6}$               & $1.63 ~\times ~10^{-4}$ & $1.17 ~\times ~10^{-4}$ & $8.85 ~\times ~10^{-5}$\\
\hline
$[^3S_1]_{\mathbf{\bar 3}}$ & $3.76 ~\times ~10^{-3}$ & $2.68 ~\times ~10^{-3}$ & $2.01 ~\times ~10^{-3}$\\
\hline
$[^1P_1]_{\mathbf{\bar 3}}$ & $2.87 ~\times ~10^{-5}$ & $2.05 ~\times ~10^{-5}$ & $1.54 ~\times ~10^{-5}$\\
\hline
$[^3P_0]_{6}$               & $3.90 ~\times ~10^{-6}$ & $2.79 ~\times ~10^{-6}$ & $2.10 ~\times ~10^{-6}$\\
\hline
$[^3P_1]_{6}$               & $3.52 ~\times ~10^{-6}$ & $2.53 ~\times ~10^{-6}$ & $1.91 ~\times ~10^{-6}$\\
\hline
$[^3P_2]_{6}$               & $1.32 ~\times ~10^{-5}$ & $9.36 ~\times ~10^{-6}$ & $7.01 ~\times ~10^{-6}$\\
\hline
total & $3.97 ~\times ~10^{-3}$ & $2.83 ~\times ~10^{-3}$ &   $2.12~\times ~10^{-3}$     \\
\hline
\end{tabular}
\label{mb_asuncer}
\end{center}
\end{table}

\section{Summary}\label{sec4}

In this paper, the photoproduction of doubly heavy baryons $\Xi_{cc}$, $\Xi_{bc}$, $\Xi_{bb}$ and their $P$-wave excited states have been investigated systematically within the framework of NRQCD. The subprocess of photoproduction for $\Xi_{QQ^{\prime}}$ is mainly via $\gamma+\gamma \rightarrow \langle QQ^{\prime} \rangle[n] +\bar{Q^{\prime}}+\bar{Q} \rightarrow \Xi_{QQ^{\prime}} +\bar{Q^{\prime}}+\bar{Q}$, where $Q^{(\prime)}$ denotes as the heavy quark $b$ or $c$, $[n]$ is the color and spin quantum number of intermediate diquark and can be $[{}^3S_1]_{\bar{\textbf{3}}/\textbf{6}}$ and $[{}^1S_0]_{\bar{\textbf{3}}/\textbf{6}}$ for $S$-wave states, $[{}^1P_1]_{\bar{\textbf{3}}/\textbf{6}}$ and
$[{}^3P_J]_{\bar{\textbf{3}}/\textbf{6}}$ with $J=0,~1,~2$ for $P$-wave states.

With the increase of collision energy $\sqrt{s}$ at CEPC and FCC-ee, the cross section for the photoproduction of $\Xi_{QQ^{\prime}}$ increases rapidly at first and then decreases, and the largest result appears in the energy range of tens of GeV. Therefore, four typical designed collision energy $\sqrt{s}=91, 160, 240, 360$ GeV are selected to do the numerical analysis. The numerical results show that at $\sqrt{s}=91$~GeV, the contribution of photoproduction for $P$-wave $\Xi_{cc}$, $\Xi_{bc}$, and $\Xi_{bb}$ is approximately $2.19\%$, $4.23\%$, $1.26\%$ of the contribution for $S$-wave, respectively. As the collision energy increases, the contribution of $P$-wave also increases, up to $2.46\%$, $4.96\%$, $1.60\%$. 

Assuming that the highly excited state can decay into ground state 100\%, both at CEPC and FCC-ee, a large number of $\Xi_{QQ^{\prime}}$ events can be produced. In particular, when $\sqrt{s} = 91$~GeV, the number of produced $\Xi_{cc}$, $\Xi_{bc}$, and $\Xi_{bb}$ events is as high as $10^8$, $10^7$ and $10^5$ respectively, which is very promising to be detected in future experiments. At the same luminosity, the ratios of $\Xi_{cc}$, $\Xi_{bc}$, and $\Xi_{bb}$ events produced at the collision energies $\sqrt{s} = 91, 160, 240, 360$ are about $863:44:1$, $591:36:1$, $496:32:1$, and $431:30:1$, respectively.

To show the dynamical behavior of the produced $\Xi_{QQ^{\prime}}$ baryon from different intermediate diquark state $\langle QQ^{\prime} \rangle[n]$ and be helpful for the experiments measurements, we present the differential cross sections, involving the angular, invariant mass, transverse momentum, and rapidity distributions, of $\Xi_{cc}$, $\Xi_{bc}$ and $\Xi_{bb}$, and theoretical uncertainty at collision energy $\sqrt{s}=91$ GeV. Two main sources of theoretical uncertainty in the photoproduction of $\Xi_{QQ^{\prime}}$ baryons are emphatically analyzed, i.e., the heavy quark masses $m_c$ and $m_b$ and the renormalization energy scale $\mu$. The dependence of the cross section from each intermediate diquark state on heavy quark mass is significant. The cross section tends to decrease with the increase of $m_c$ and $m_b$ mainly due to the inhibition of phase space. The dependence of $\Xi_{cc}$ baryon on $m_c$ is obviously greater than that of $\Xi_{bc}$ baryon. Similarly, the $\Xi_{bb}$ baryon is significantly more dependent on $m_b$ than $\Xi_{bc}$ baryon. We numerically analyze the theoretical uncertainty arising from heavy quark mass by summing the contributions of all intermediate diquark states (total). Specifically, the total cross section of $\Xi_{cc}$ baryon can vary by $-53.46\%\sim +145.31\%$ for $m_c=1.8\pm 0.3$ GeV; the total cross section of $\Xi_{bc}$ baryons can vary by $-31.53\%\sim +57.26\%$ for $m_c=1.8\pm 0.3$ and $-16.13\%\sim+18.55\%$ for $m_b=5.1\pm 0.3$ GeV; and the total cross section of $\Xi_{bb}$ baryons can vary by $-27.82\%\sim+39.79\%$ for $m_b=5.1\pm 0.3$ GeV.

By changing the energy scale $\mu=\frac{1}{2}\mu_0$ and 2$\mu_0$, the total cross section will change by $-30.17\%\sim+67.36\%$, $-24.03\%\sim+54.03\%$, and $-25.09\%\sim+40.28\%$
for the photoproduction of $\Xi_{cc}$, $\Xi_{bc}$, and $\Xi_{bb}$, respectively. With the help of subsequent higher-order calculations, the energy scale uncertainty will be greatly reduced until more accurate numerical results are obtained. Nevertheless, the photoproduced events of doubly heavy baryons $\Xi_{QQ^{\prime}}$ remain appreciable at CEPC and FCC-ee, and further experimental testing of the doubly heavy baryons in future will be expected.

 {\bf Acknowledgments:} 
Thanks to Xi-Jie Zhan for a very helpful discussion. This work was partially supported by the Natural Science Foundation of Guangxi (no. 2024GXNSFBA010368, 2024JJA110093). This work was also supported by the Central Government Guidance Funds for Local Scientific and Technological Development, China (no. Guike ZY22096024) and the National Natural Science Foundation of China (no. 12005045).


\end{document}